\documentclass[12pt]{iopart}
\usepackage{amssymb}
\usepackage{amsfonts}
\usepackage{psfrag}
\usepackage{graphicx}
\usepackage{color}
\usepackage{cite}
\eqnobysec

\def\keywords#1{\vspace{10pt}
     \begin{indented}
     \item[]\rm Keywords: #1\par
     \end{indented}}

\def\be{\begin{equation}}
\def\ee{\end{equation}}
\def\bea{\begin{eqnarray}}
\def\eea{\end{eqnarray}}

\def\CL{{\mathcal L}}

\def\CT{{\mathcal T}}
\def\CS{{\mathcal S}}

\def\fT{{\mathfrak T}}

\def\fS{{\mathfrak S}}

\def\scal{\stackrel{\mathrm{s.l.}}{=}}

\def\dseg{D_{\mathrm{seg}}}
\begin{document}
\jl{1}

\title[Two-species diffusion-annihilation process]{Two-species diffusion-annihilation process 
on the fully-connected lattice: probability distributions and extreme value statistics}
\author{Lo\"\i c Turban}

\address{Laboratoire de Physique et Chimie Th\'eoriques, Universit\'e de Lorraine--CNRS (UMR7019),
Vand\oe{u}vre l\`es Nancy Cedex, F-54506, France} 

\ead{loic.turban@univ-lorraine.fr}

\begin{abstract} We study the two-species diffusion-annihilation process, $A+B\rightarrow$~\O,
on the fully-connected lattice. Probability distributions for the number of particles and 
the reaction time are obtained for a finite-size system using a master equation approach. 
Mean values and variances are deduced from generating functions. When the reaction is far 
from complete, i.e. for a large number of particles of each species, mean-field theory is exact
and the fluctuations are Gaussian. 
In the scaling limit the reaction time displays extreme-value statistics in the vicinity of 
the absorbing states. A generalized Gumbel distribution is obtained for unequal initial densities, 
$\rho_A>\rho_B$. For equal or almost equal initial densities, 
$\rho_A\simeq\rho_B$, the fluctuations of the reaction time near the absorbing state 
are governed by a probability density involving derivatives of $\vartheta_4$, 
the Jacobi theta function.
\end{abstract}


\submitto{J. Phys. A: Mathematical and Theoretical}

\section{Introduction}

In the field of non-equilibrium statistical mechanics, reaction-diffusion processes 
offer the possibility to study the effects of fluctuations on conceptually very 
simple model systems like the single-species or the two-species annihilation 
processes~\cite{alcaraz94,hinrichsen00,benavraham00,schutz01,odor04,henkel08,odor08,krapivsky10a,tauber17}.

In a standard mean-field approximation~\cite{smoluchowski16}, the bimolecular reaction 
$A+B\rightarrow$~\O\ displays a $t^{-1}$ asymptotic decay of the particle densities 
for equal initial values, $\rho_A(0)=\rho_B(0)$. For unequal densities, $\rho_A>\rho_B$, 
the approach to the 
absorbing state, $\rho_B=0$ and $\rho_A=\rho_A(0)-\rho_B(0)$, is exponential.
The mean-field approximation assumes that the system remains homogeneous and ignores 
the effect of spatial correlations in the distribution of reactants, thus giving a lower bound to
the actual particle densities~\cite{burlatsky87}.

The relevance in low dimensions of initial concentration fluctuations was pointed out
by Ovchinnikov and Zeldovich~\cite{ovchinnikov78} who found a $t^{-3/4}$ decay in dimension $D=3$
for equal initial densities.
This result was soon generalized and a $t^{-D/4}$ 
decay was proposed for $\rho_A(0)=\rho_B(0)$ on the basis of numerical simulations, approximate 
analytical approaches and scaling 
arguments~\cite{toussaint83,kang84}~\footnote[1]{
Note that for initially separated reactants the kinetics is inhomgeneous and governed 
by the reaction in a growing domain around the interface~\cite{galfi88,krapivsky10b}.}. 
The validity of this asymptotic behaviour was later confirmed by establishing
rigorous bounds on the particle density~\cite{bramson88,bramson91} and 
through a renormalization group study~\cite{lee95,tauber05}. 

The slowing down of the process is due to the segregation of $A$ and $B$ particles
into $A$-rich and $B$-rich domains, at the scale of the diffusion 
length~\cite{kang84,bramson91,leyvraz92}.
The segregation is a consequence of the initial fluctuations of the densities around their mean values.
At long time the reaction is efficient only at the interface between the domains and thus slows down.
This effect is relevant below the segregation dimension $\dseg=4$ 
at which the $t^{-1}$ homogeneous mean-field decay is recovered. 
When generalized to $q$ species~\cite{benavraham86} the problem has a segregation dimension
$\dseg(q)=4/(q-1)\geq2$~\cite{hilhorst04}.

When $\rho_A(0)>\rho_B(0)$ the density of the minority reactant behaves asymptotically as 
\be
\rho_B(t)\sim\e^{-\lambda_DG_D(t)}\,,
\label{rhobgdt}
\ee
with~\cite{bramson88,bramson91}
\be
G_D(t)=\left\{
\begin{array}{ll}
\sqrt{t}\,,& D=1\\
t/\ln t\,,& D=2\\
t\,,& D=3
\end{array}
\right.
\label{gdt}
\ee
Note that the upper critical dimension, 
as for the single-species process, is $D_{\mathrm c}=2$~\cite{hilhorst04,tauber05}. The 
$t^{-D/4}$ behaviour can be actually obtained using mean-field rate equations, provided the 
inhomogeneity of the system is taken into account. Although there is no qualitative change
at $D_{\mathrm c}$ for equal initial densities, the upper critical dimension signals itself
via logarithmic corrections at $D_{\mathrm c}$ and a stretched exponential decay below 
$D_{\mathrm c}$ for unequal initial densities.

The two-species annihilation process has potential applications in different domains. It can be 
used to model particle-antiparticle 
annihilation in the early universe~\cite{toussaint83,lee84}, the kinetics of
bimolecular chemical reactions~\cite{ovchinnikov89,savara10} or 
electron-hole recombination in irradiated semiconductors~\cite{vardeny80}.

The aim of the present work is to study analytically the kinetics 
of the two-species reaction-diffusion process 
on the fully-connected lattice with an emphasis on probability distributions. 
This is a continuation of previous work on the single-species 
process~\cite{turban18}. Since the lattice with $N$ sites can only be embedded in a $(N-1)$-dimensional 
space, taking the thermodynamic limit requires an infinite-dimensional space and one expects mean-field 
behaviour. Our purpose is to obtain exact results for the particle density and the 
reaction time in finite-size systems and to study the 
extreme-value statistics of the reaction time, in the vicinity of the absorbing state, 
for both equal or unequal initial densities of the reactants.

The paper is organized as follows. In section 2 we present the model, its mean-field solution when 
homogeneity is assumed and give a brief description of our results. In section 3
we study the statistics of the number of particles surviving at a given time, first on a finite system 
and then in the scaling limit. Section 4 is devoted to a similar 
study of the reaction time, i.e. the time 
needed to have a given number of particles remaining. This is followed by the conclusion in section 5. 
Details of the calculations are given in six appendices.

\section{Model, mean field and main results}

\subsection{Model}

\begin{table}
\caption{\label{t-1} $s_A$ and $s_B$ are the particle numbers, $\rho_A$ 
and $\rho_B$ their densities, $N$ is the number of sites and $n=N/2$. The last line gives the 
relations for mean values and variances.}
\begin{indented}
\item[]\begin{tabular}{@{}llll}
\br
$s_A=s+d$&$s_B=s-d$&$s=\frac{s_A+s_B}{2}$&$d=\frac{s_A-s_B}{2}$\\
\ms
$\rho_A=\frac{s_A}{N}$&$\rho_B=\frac{s_B}{N}$&$x=\frac{s}{n}$&$y=\frac{d}{n}$\\
\ms
$\rho_A=\frac{x+y}{2}$&$\rho_B=\frac{x-y}{2}$&$x=\rho_A+\rho_B$&$y=\rho_A-\rho_B$\\
\ms
$s_A-\overline{s_A}=s-\overline{s}$&$s_B-\overline{s_B}=s-\overline{s}$&
$\overline{\Delta s_A^2}=\overline{\Delta s^2}$&$\overline{\Delta s_B^2}=\overline{\Delta s^2}$\\
\br
\end{tabular}
\end{indented}
\end{table}

We consider the two-species reaction-diffusion process, $A+B\rightarrow$~\O, 
on a fully connected lattice with $N$ sites. 
Let $s_A$ and $s_B$ be the number of particles of each type with $s_A\geq s_B$ 
and at most one particle per site. In the following we shall use the variables
\be
s=\frac{s_A+s_B}{2}\,,\qquad d=\frac{s_A-s_B}{2}\,.
\label{sd}
\ee
Thus $d=0$ when the initial densities are equal and $s=d$ when the reaction is complete. 
A dictionary giving the relations with standard notations is given in table~\ref{t-1}.

The system evolves in time through random sequential updates. 
An update consists of one or two steps. A first site $i$ is selected at random among the $N$.  
When this site is occupied by a particle of type $A$ ($B$) a second site $j$ is randomly 
selected among the $N$. If the destination site is occupied by a particle of type $B$ ($A$), 
the two particles annihilate and $s\rightarrow s-1$ . In all other cases $s$ is unchanged. 
$d$ is always conserved. At each update the time $t$ is incremented 
by $1/N$ so that $t=k/N$ where $k$ is the number of updates. Note that first selecting a site
instead of a particle is vital to keep a constant time increment.

The probabilities for the different events are the following:
\begin{itemize}
\item $s\rightarrow s'=s-1$, with probability:
\be\fl
2\,\frac{s^2-d^2}{N^2}=
\underbrace{s_A/N}_{i=A}\times
\underbrace{s_B/N}_{j=B}\,
+\,\underbrace{s_B/N}_{i=B}\times
\underbrace{s_A/N}_{j=A}
\label{probrea}
\ee
\item $s\rightarrow s'=s$, with probability:
\be\fl
1-2\,\frac{s^2-d^2}{N^2}=\underbrace{1-2s/N}_{i=\emptyset}\,+\,
\underbrace{s_A/N}_{i=A}\times
\underbrace{1-s_B/N}_{j\not=B}
\,+\,\underbrace{s_B/N}_{i=B}\times
\underbrace{1-s_A/N}_{j\not=A}
\label{probnorea}
\ee
\end{itemize}
Note that, contrary to what occurs on finite-dimensional lattices, 
the initial distribution of the two species does not matter for the fully-connected lattice.

\subsection{Mean field solution}
In the following we always assume that in the initial state, at $t=0$, all the sites are occupied, 
$s_A+s_B=N$ and we neglect the effect of spatial fluctuations ($D>\dseg$). The initial value of $s$ is then $n=N/2$ according to~\eref{sd}. 
In the scaling limit (s.l.), when $N$ and $n\to\infty$, we introduce the scaled variables
\be 
x\scal\frac{s}{n}=\rho_A+\rho_B\,,\qquad 
y\scal\frac{d}{n}=\rho_A-\rho_B\,.
\label{xy}
\ee
where $\rho_A$ and $\rho_B$ are the particle densities.
Note that $x$ is the fraction of occupied sites at $t$ and $y$ 
is a constant giving the asymptotic value of this fraction when $t\to\infty$.

After a small number of updates, $\Delta k$, according to~\eref{probrea} 
the mean value of $s$ is changed by 
\be 
\Delta s=-2\,\frac{s^2-d^2}{N^2}\Delta k\,
\label{deltas}
\ee
where on the right the fluctuations of $s$ 
around its mean value are neglected. In the scaling limit,
$\Delta s\to n\,dx$, $\Delta k\to N\,dt=2n\,dt$, yielding
\be
\frac{dx}{dt}=-\frac{s^2-d^2}{n^2}=-(x^2-y^2)\,,
\label{edif-1}
\ee
so that
\be
\frac{dx}{x^2-y^2}=\frac{1}{2y}\left(\frac{dx}{x-y}-\frac{dx}{x+y}\right)=-dt\,.
\label{edif-2}
\ee
Finally the solution satisfying the initial condition, $x=1$ when $t=0$, is given by:
\be
x=y\,\frac{1+y+(1-y)\,\e^{-2yt}}{1+y-(1-y)\,\e^{-2yt}}\,.
\label{xmf}
\ee
This yields
\bea
\rho_A&=\frac{s_A}{N}=\frac{x+y}{2}=\frac{y(1+y)}{1+y-(1-y)\,\e^{-2yt}}\,,\nonumber\\
\rho_B&=\frac{s_B}{N}=\frac{x-y}{2}=\frac{y(1-y)\,\e^{-2yt}}{1+y-(1-y)\,\e^{-2yt}}\,,
\label{rhoArhoB-1}
\eea
for the densities of the two species. When $y>0$ the approach to the asymptotic values, 
$\rho_A=y$ and $\rho_B=0$, is exponential. When $y\to0$ an algebraic decay is obtained:
\be
\rho_A=\rho_B=\frac{1}{2(t+1)}\,.
\label{rhoArhoB-2}
\ee
\begin{figure}[!th]
\begin{center}
\includegraphics[width=8cm,angle=0]{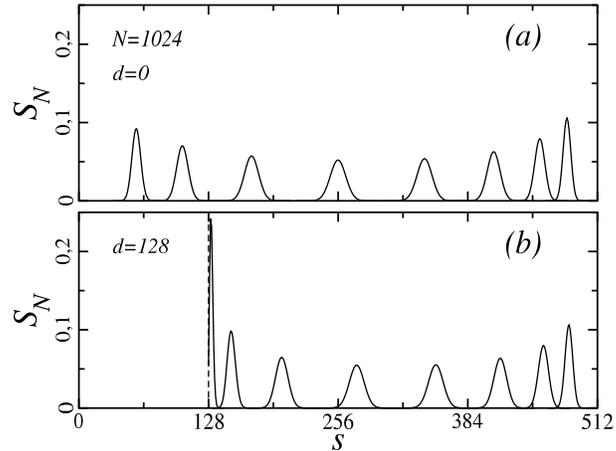}
\end{center}
\vglue -.5cm
\caption{Evolution of the probability distribution of $s$, $S_N(s,k)$, 
for values of the number of updates $k=2^i$, $i=6,\ldots,13$ from right to 
left on a lattice with $N=1024$ sites. The reaction is complete 
when (a) $s=d=0$ , (b) $s=d=128$. The 
fluctuations are Gaussian in the scaling limit and maximum for $k$ close to $N$.   
\label{fig-1}
}
\end{figure}
\begin{figure}[!th]
\begin{center}
\includegraphics[width=8cm,angle=0]{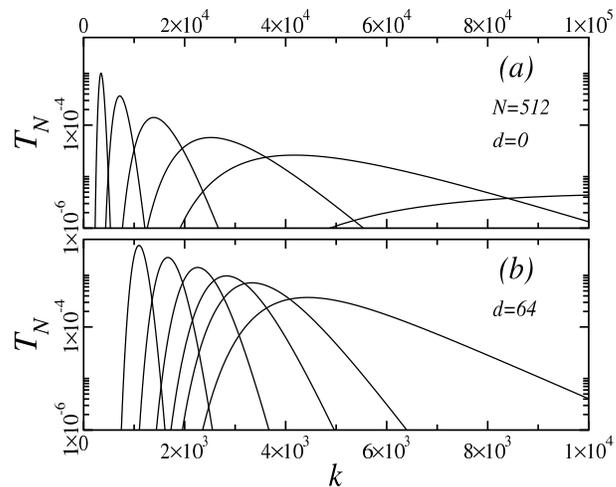}
\end{center}
\vglue -.5cm
\caption{Semi-logarithmic plot of the probability distribution $T_N(s,k)$ 
of the number of updates $k$ needed to reach some given values of $s=d$, $d+2^i$, 
$i=1,\ldots,5$ from right to left, for (a) $d=0$, (b) $d=64$ on a lattice with 
$N=512$ sites. The reaction time is given by $k/N$.
The probability distribution evolves from a Gaussian behaviour when $s\gg d$ to a wide 
asymmetric extreme-value distribution when the system approaches its absorbing state.
Note that the two $k$ scales differ by a factor of 10.
\label{fig-2}
}
\end{figure}
The time needed to reach a given value of $x$ is
\be
t=\frac{1}{2y}\ln\left[\frac{(1-y)(x+y)}{(1+y)(x-y)}\right]\,,
\label{tmf-1}
\ee
leading to
\be
t=\frac{1}{x}-1\,,
\label{tmf-2}
\ee
when $y\to0$.

\subsection{Main results}

The evolution of the system is illustrated in figures~\ref{fig-1} and~\ref{fig-2} 
giving respectively the probability distribution $S_N(s,k)$ of $s$ at different reaction times and 
the probability distribution $T_N(s,k)$ of the reaction time at different values of $s$.

As expected for an infinite-dimensional system the mean value of $s$ at time $t$ 
is in agreement with mean-field theory. Asymptotically, it decays as 
$t^{-1}$ when $s_A=s_B$ and approaches its asymptotic value exponentially when $s_A>s_B$. 
The mean value and the variance both scale as $n$ and the fluctuations of $s$ are Gaussian.

For the statistics of the reaction time $t$ three different 
regimes are observed with the following results in the scaling limit: 

\begin{itemize}
\item $x>y\geq0$:~The reaction is far from complete with $s_A=\Or(n)$, $s_B=\Or(n)$ and $s_A\geq s_B$. 
The mean values of the reaction time are the mean-field ones:
\be\fl
\overline{t_N}=\frac{1}{x}-1\,,\quad y=0\,;\qquad
\overline{t_N}=\frac{1}{2y}\ln\left[\frac{(1-y)(x+y)}{(1+y)(x-y)}\right]\,,\quad y>0\,.
\label{x>y-1}
\ee
The fluctuations are weak, the variance scaling as $n^{-1}$:
\bea
\fl\overline{\Delta t_N^2}&=\frac{\chi_y(x)}{n}\,;\qquad
\chi_0(x)=\frac{1}{3x^3}-\frac{1}{2x}+\frac{1}{6}\,,\quad y=0\,;\nonumber\\
\fl\chi_y(x)&=\frac{1}{2y^2}\!\left(\frac{x}{x^2\!-\!y^2}-\frac{1}{1\!-\!y^2}\right)\!
-\!\frac{1}{4y}\left(1\!+\!\frac{1}{y^2}\right)
\ln\left[\frac{(1\!-\!y)(x\!+\!y)}{(1\!+\!y)(x\!-\!y)}\right]\,,\quad y>0\,.
\label{x>y-2}
\eea
The probability density is Gaussian:
\be
\fT(x,\theta)=\frac{\e^{-\theta^2/[2\chi_y(x)]}}{\sqrt{2\pi\chi_y(x)}}\,,\quad
\theta=n^{1/2}(t-\overline{t_N})\,.
\label{x>y-3}
\ee
\item $x=y>0$, $s=ny+u$:~ 
The reaction is close to completion with unequal numbers of  
particles ($s_A-s_B=\Or(n)$ and $u=s_B=\Or(1)$). The reaction time scales logarithmically with $n$
\be 
\overline{t_N}=\frac{1}{2y}\left[\ln\left(2ny\frac{1-y}{1+y}\right)
+\gamma-H_u\right]\,,
\label{x=y-1}
\ee
where $H_u=\sum_{j=1}^u1/j$ is a harmonic number and  $\gamma=0.577215665\ldots$ is the Euler constant. 
The variance is independent of $n$:
\be
\overline{\Delta t_N^2}=\frac{1}{4y^2}\left[\zeta(2)-H_u^{(2)}\right]\,.
\label{x=y-2}
\ee
Here $H_u^{(2)}$ is a generalized harmonic number such that $H_l^{(m)}=\sum_{j=1}^l1/j^m$
and $\zeta(2)=\pi^2/6$.
The system displays extreme value statistics. The fluctuations are governed by a generalized Gumbel 
distribution~\cite{ojo2001,pinheiro2016}, indexed by $u$:
\be\fl
\fT'(u,\theta')=\frac{1}{u!}\exp\left[-(u+1)(\theta'+\gamma-H_u)
-\e^{-(\theta'+\gamma-H_u)}\right]\,,\quad\theta'=2y(t-\overline{t_N})\,.
\label{x=y-3}
\ee

\item $x=y=0$, $s\geq d$: 
~The reaction is close to completion with $s_A=\Or(1)\geq s_B\geq0$.
The reaction time grows as $n$
\be\fl
\overline{t_N}=n(\zeta(2)-H_s^{(2)})\,,\quad d=0\,;\qquad
\overline{t_N}=\frac{n}{2d}(H_{s+d}-H_{s-d})\,,\quad d>0\,,
\label{x=y=0-1}
\ee
and the variance as $n^2$
\bea
\overline{\Delta t_N^2}&=n^2(\zeta(4)-H_s^{(4)})\,,\quad d=0\,;\nonumber\\
\overline{\Delta t_N^2}&=\frac{n^2}{4d^2}
\left[2\zeta(2)\!-\!H_{s-d}^{(2)}\!-\!H_{s+d}^{(2)}\!+\!\frac{1}{d}(H_{s-d}\!-\!H_{s+d})\right]
\,,\quad d>0\,,
\label{x=y=0-2}
\eea
where $\zeta(4)=\pi^4/90$. The fluctuations of the reaction time are even stronger 
and  governed now by derivatives of the Jacobi theta function $\vartheta_4$:
\be\fl
\fT''(s,\theta'')=\frac{\e^{d^2\theta''}}{(s-d)!(s+d)!}
\prod_{m=0}^s(m^2+d/d\theta'')\,\vartheta_4\left(0,\e^{-\theta''}\right)\,,\quad
\theta''=\frac{t}{n}\,.
\label{x=y=0-3}
\ee
\end{itemize}

\section{Number of surviving particles at a given time}
In this section we study the probability distribution $S_N(s,k)$ giving the probability to have $s+d$ 
particles of type $A$ and $s-d$ particles of type $B$ remaining after $k$ updates. As above we assume 
that the $N$ sites are initially occupied, $s_A+s_B=N$.

\subsection{Master equation}
According to \eref{probrea} and \eref{probnorea} the master equation governing the evolution of 
the system takes the following form
\be\fl
S_N(s,k)\!=\!\left(1\!-\!2\,\frac{s^2\!-\!d^2}{N^2}\right)\!S_N(s,k\!-\!1)\!
+\!2\,\frac{(s\!+\!1)^2\!-\!d^2}{N^2}\,S_N(s\!+\!1,k\!-\!1)\,,\quad s=d,\ldots,n\,,
\label{master}
\ee
with the boundary condition $S_N(s>n,k)=0$ and the initial condition $S_N(s,0)=\delta_{s,n}$.
In~\eref{master} the first (second) term on the right gives the probability to 
be in a state with $2s$ particles ($2s+2$ particles) after $k-1$ updates and to remain in 
this state (to have two particles annihilating) at the $k$th update.

\subsection{Eigenvalue problem}
Let us define the column state vector $|S_N(k)\rangle$ with components $S_N(s,k),\quad s=d,\ldots,n$,
the master equation~\eref{master} can be written in matrix form as 
$|S_N(k)\rangle=\mathsf{T}|S_N(k-1)\rangle$ where the transition matrix $\mathsf{T}$ is given by:
\be\fl
\mathsf{T}=\left(\begin{array}{cccccc}
1       &2\,\frac{(d+1)^2-d^2}{N^2}  &0	            &0                   &0                 &0                   \\
0       &1\!-\!2\,\frac{(d+1)^2-d^2}{N^2}&2\,\frac{(d+2)^2-d^2}{N^2} &0                   &0                 &0                   \\
	&                      &\ddots              &\ddots              &                  &                    \\
0	&0                     &0                   &1\!-\!2\,\frac{s^2-d^2}{N^2}&2\,\frac{(s+1)^2-d^2}{N^2}&0                   \\
        &                      &                    &                    &\ddots            &\ddots              \\ 
0	&0	               &0	            & 0	                 &0                 &1\!-\!2\,\frac{n^2-d^2}{N^2}\\
\end{array}\right)\,.
\label{tmatrix}
\ee
The eigenvalue equation $\mathsf{T}|v^{(r)}\rangle=\lambda_r|v^{(r)}\rangle$ leads to the linear system
\be
\left(1-2\,\frac{s^2-d^2}{N^2}-\lambda_r\right)v_s^{(r)}+2\,
\frac{(s\!+\!1)^2-d^2}{N^2}\,v_{s+1}^{(r)}=0\,,
\quad s=d,\ldots,n\,,
\label{lin}
\ee
with $v_{n+1}^{(r)}=0$. It is easy to verify that 
\be\fl
\lambda_r=1-2\,\frac{r^2-d^2}{N^2}\,,\qquad 
v_s^{(r)}=\left\{
\begin{array}{ccc}
 (-1)^{r-s}
 v_r^{(r)}\prod_{j=1}^{r-s}\frac{(s+j)^2-d^2}{r^2-(s+j-1)^2}&\mathrm{when}&s<r\\
 \ms
 0&\mathrm{when}&s>r
\end{array}
\right.\!,
\label{sol}
\ee
solves the eigenvalue problem~\eref{lin}. The solution involves the repeated use of the recursion relation
\be
v_s^{(r)}=-\frac{(s+1)^2-d^2}{r^2-s^2}\,v_{s+1}^{(r)}\,,
\label{rec}
\ee
which follows from~\eref{lin}.
The value of $v_r^{(r)}$, which remains free, will be used 
to satisfy the initial condition.

\subsection{Probability distribution $S_N(s,k)$}
We look for the initial state vector under the form $|S_N(0)\rangle=\sum_{r=d}^n|v^{(r)}\rangle$ 
which leads to the condition
\be
S_N(s,0)=\sum_{r=s}^nv_s^{(r)}=\delta_{s,n}\,,\qquad s=d,\ldots,n\,,
\label{sns0}
\ee
for the components. From the values of $v_r^{(r)}$ with $r=n,\ldots,n-3$ (see appendix A) 
we can infer that the general expression reads:
\be
v_r^{(r)}=\frac{(n-d)!(n+d)!}{(n-r)!(n+r)!}{2r\choose r-d}\,,\qquad r=d,\ldots,n\,.
\label{vrr}
\ee
Then, according to~\eref{sol}, one obtains:
\be
v_s^{(r)}=\frac{(-1)^{r-s}2r(n-d)!(n+d)!(r+s-1)!}{(n-r)!(n+r)!(s-d)!(s+d)!(r-s)!}
\,,\quad s\leq r\,.
\label{vsr}
\ee
After $k$ updates the state vector $|S_N(k)\rangle$ is given by
\be
\mathsf{T}^k|S_N(0)\rangle=\sum_{r=d}^n\mathsf{T}^k|v^{(r)}\rangle
=\sum_{r=d}^n\lambda_r^k|v^{(r)}\rangle\,,
\label{tksn0}
\ee
which, according to~\eref{sol} and~\eref{vsr}, gives
\be
\fl S_N(s,k)\!=\!\sum_{r=s}^n\lambda_r^kv_s^{(r)}\!
=\!\sum_{r=s}^n
\frac{(-1)^{r-s}2r(n-d)!(n+d)!(r+s-1)!}{(n-r)!(n+r)!(s-d)!(s+d)!(r-s)!}
\left(\!1\!-\!2\,\frac{r^2\!-\!d^2}{N^2}\right)^k
\label{snsk}
\ee
for the components.

\subsection{Mean value and variance when $d=0$}

\begin{figure}[!th]
\begin{center}
\includegraphics[width=8cm,angle=0]{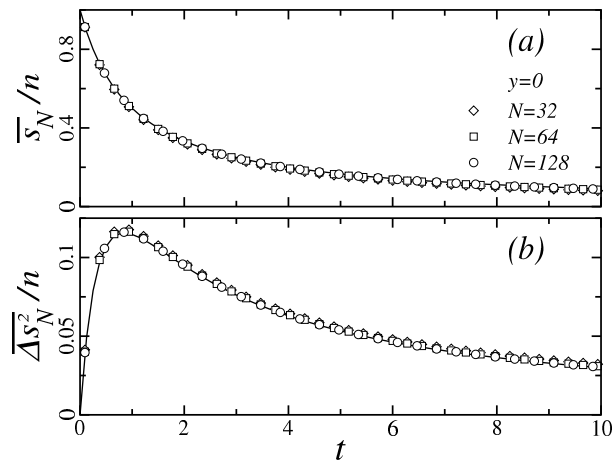}
\end{center}
\vglue -.5cm
\caption{
\label{fig-3} Scaling behaviour of (a) the mean value $\overline{s_N}$ and 
(b) the variance $\overline{\Delta s_N^2}$ of $s=(s_A+s_B)/2$ as a function of the time $t=k/N$.
In the initial state $\rho_A=\rho_B=1/2$. The finite-size data for $N=32$ 
(diamond), 64 (square), 128 (circle) were deduced from $S_N(s,k)$ given by a numerical 
iteration of the master equation~\eref{master}. A good collapse on 
the full lines corresponding to the scaling functions in~\eref{snt} and~\eref{dsn2t} is obtained.
The variance is maximum for a number of updates $k$ close to $N$. 
}
\end{figure}

Let us define the generating function
\be\fl 
\CS_N(w,k)=\sum_{s=d}^nsw^sS_N(s,k)=\sum_{r=d}^n\frac{r(n-d)!(n+d)!}{(n-r)!(n+r)!}
\left(\!1\!-\!2\,\frac{r^2\!-\!d^2}{N^2}\right)^k\Omega_{r,d}(w)
\label{snwk}
\ee
where
\be
\Omega_{r,d}(w)=\sum_{s=d}^r(-1)^{r-s}{r+s\choose s-d}{r+d\choose s+d}
\frac{2sw^s}{r+s}\,.
\label{ordw}
\ee
In appendix B we show that when $d=0$
\be
\Omega_{r,0}(1)=2
\label{or01-1}
\ee
and
\be
\left.\frac{d\Omega_{r,0}}{d w}\right|_{w=1}=2r^2\,,
\label{dor01-1}
\ee
which allows us to evaluate the mean value of $s$
\be
\overline{s_N(k)}=\CS_N(1,k)=\sum_{r=1}^n2r\prod_{j=1}^r\frac{n-j+1}{n+j}
\left(\!1\!-\!2\,\frac{r^2}{N^2}\right)^k\,,\quad d=0\,,
\label{snk}
\ee
and its mean-square value:
\be
\overline{s_N^2(k)}=\left.\frac{\partial\CS_N}{\partial w}\right|_{w=1}
\!\!\!\!\!=\sum_{r=1}^n2r^3\prod_{j=1}^r\frac{n-j+1}{n+j}
\left(\!1\!-\!2\,\frac{r^2}{N^2}\right)^k\,,\quad d=0\,.
\label{sn2k}
\ee
In the scaling limit ($N,n,k\to\infty$, $t=k/N$) studied in appendix C, one obtains:
\be\fl
\frac{\overline{s_N(t)}}{n}\scal\frac{1}{t+1}-\frac{1}{6n}\left(1-\frac{1}{(t+1)^3}\right)\,,
\quad\frac{\overline{s_N^2(t)}}{n^2}\scal\frac{1}{(t+1)^2}-\frac{t}{2n(t+1)^4}\,,\quad d=0\,.
\label{snt}
\ee
In these expressions we kept the sub-leading contributions since the leading ones 
vanish in the variance given by:
\be
\frac{\overline{\Delta s_N^2(t)}}{n}\scal\frac{1}{3(t+1)}-\frac{1}{2(t+1)^3}+\frac{1}{6(t+1)^4}\,,
\qquad d=0\,.
\label{dsn2t}
\ee
Thus the fluctuations are small and $s$ is self-averaging. A comparison with finite-size data
is shown in figure~\ref{fig-3}.
We were not able to evaluate $\Omega_{r,d}(w)$ when $d>0$. This case is treated
directly in the scaling limit in the next section.

\subsection{Scaling limit when $d\geq0$}
\label{sec-scalsn}
\begin{figure}[!th]
\begin{center}
\includegraphics[width=8cm,angle=0]{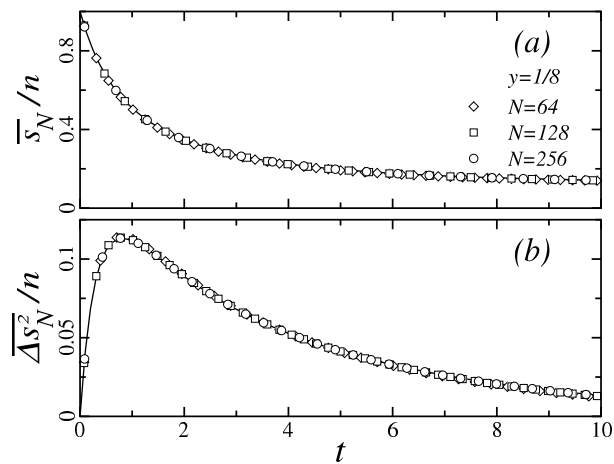}
\end{center}
\vglue -.5cm
\caption{
\label{fig-4} As in figure \ref{fig-3} for $y=\rho_A-\rho_B=1/8$ 
and $N=64$ (diamond), 128 (square), 256 (circle). The scaling functions
(full lines) are given in~\eref{xmean} and~\eref{kappat-1}.
}
\end{figure}

Let us assume that in the scaling limit, for any value of $d$, $\overline{s_N}$ 
and $\overline{\Delta s_N^2}$ are both growing as $n$ as 
in~\eref{snt} and~\eref{dsn2t} for $d=0$. This suggests the introduction, 
besides the time variable $t=k/(2n)$ and the density 
$y=d/n$, of the scaled and centered variable
\be
\sigma(s,k)=\frac{s-\overline{s_N(k)}}{n^{1/2}}\,.
\label{sigma}
\ee
Furthermore let us write the unknown mean value $\overline{x}$ as
\be
\overline{x}=\frac{\overline{s_N}}{n}=g_y(t)\,,
\label{g}
\ee
and define the probability density:
\be
\fS(\sigma,t)\scal n^{1/2}S_N(s,k)\,.
\label{ssigmat-1}
\ee
Starting  from the master equation~\eref{master} with $S_N$ replaced by $\fS$, 
a Taylor expansion of the right-hand-side
up to second order in $s$ and $k$, when re-expressed in terms of the scaled variables, 
takes the form of an expansion in powers of $n^{-1/2}$. The terms independent of $n$ 
cancel. The terms of order $n^{-1/2}$ leads to the differential equation
\be
\frac{dg_y}{dt}=-(g_y^2-y^2)\,,
\label{dgdt}
\ee
which is the mean-field equation~\eref{edif-1} so that, according to~\eref{xmf},
\be
\overline{x}=g_y(t)=y\,\frac{1+y+(1-y)\,\e^{-2yt}}{1+y-(1-y)\,\e^{-2yt}}\,,
\label{xmean}
\ee
in agreement with~\eref{snt} when $y\to0$.
To the next order, $n^{-1}$, one obtains the following partial differential equation:
\be
\frac{\partial\fS}{\partial t}=-\frac{1}{2}\left[\frac{dg_y}{dt}
+\frac{1}{2}\left(\frac{dg_y}{dt}\right)^2\right]\frac{\partial^2\fS}{\partial\sigma^2}
+2g_y\left(\fS+\sigma\frac{\partial\fS}{\partial\sigma}\right)\,.
\label{dsdt}
\ee
\begin{figure}[!th]
\begin{center}
\includegraphics[width=8cm,angle=0]{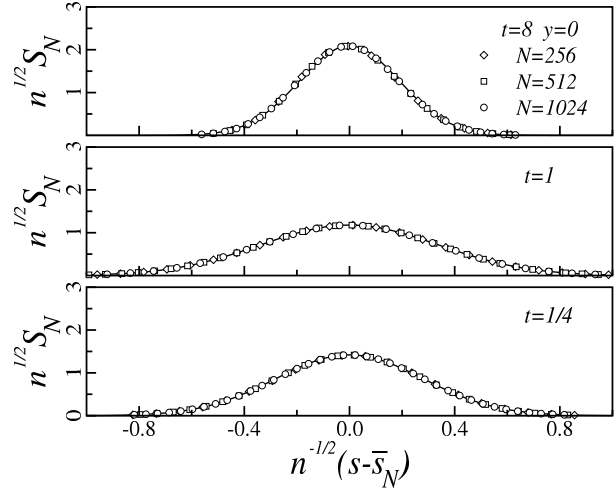}
\end{center}
\vglue -.5cm
\caption{Data collapse for the scaled probability distribution $n^{1/2}S_N(s,k)$ as 
a function of $\sigma=n^{-1/2}(s-\overline{s_N(k)})$ at different times 
$t$ and for increasing lattice sizes, $N=256$ (diamond), 512 (square) and 1024 (circle).
In the initial state y=0 so that $\rho_A=\rho_B=1/2$. The finite-size data follow from a numerical 
iteration of the master equation~\eref{master}. The full lines correspond to the 
Gaussian density~\eref{ssigmat-2} obtained in the scaling limit.
The fluctuations are stronger for $t\simeq1$.
\label{fig-5}
}
\end{figure}
Introducing the reduced variance $\kappa_y(t)=\overline{\Delta s_N^2}/n$ and assuming that 
$\fS$ depends on $t$ only through $\kappa_y$, the partial differential equation~\eref{dsdt} 
can be rewritten in the following form:
\bea\fl
\left(\frac{\partial\fS}{\partial\kappa_y}\!-\!\frac{1}{2}\frac{\partial^2\fS}{\partial\sigma^2}\right)
\!\frac{d\kappa_y}{dt}\!&=\!-\frac{1}{2}\!\left[\frac{d\kappa_y}{dt}\!+\!4g_y\kappa_y\!+\!\frac{dg_y}{dt}
\!+\!\frac{1}{2}\left(\!\frac{dg_y}{dt}\!\right)^2\right]\!\frac{\partial^2\fS}{\partial\sigma^2}
\nonumber\\
&\ \ \ \ \ \ \ \ \ \ \ \ \ \ \ \ \ \ \ \ \ \ \ \ \ \ \ \ \ \ \ \ \ \ \ \ \ \ \ 
+\!2g_y\!\left(\!\fS\!+\!\sigma\frac{\partial\fS}{\partial\sigma}
\!+\!\kappa_y\frac{\partial^2\fS}{\partial\sigma^2}\right)\!\!.
\label{dsdkap}
\eea
The left-hand-side and the last bracket on the right-hand-side are both vanishing 
for the Gaussian density with variance $\kappa_y(t)$, thus we have
\be
\fS(\sigma,t)=\frac{\e^{-\sigma^2/[2\kappa_y(t)]}}{\sqrt{2\pi\kappa_y(t)}}\,,
\label{ssigmat-2}
\ee
when, according to~\eref{dsdkap}, $\kappa_y(t)$ satisfies the first-order differential equation:
\be
\frac{d\kappa_y}{dt}+4g_y\kappa_y+\frac{dg_y}{dt}+\frac{1}{2}\left(\frac{dg_y}{dt}\right)^2=0\,.
\label{dkapdt}
\ee
The solution is discussed in appendix D and reads:
\be\fl
\kappa_y(t)\scal\frac{\overline{\Delta s_N^2}}{n}=\left[(1\!+\!y)^2
\e^{2yt}\!-\!(1\!-\!y)^2\e^{\!-\!2yt}\!-\!4y(1\!+\!t\!-\!y^4t)\right]
\frac{2y(1\!-\!y^2)\e^{\!-\!4yt}}{\left[1\!+\!y\!-\!(1\!-\!y)\e^{\!-\!2yt}\right]^4}\,.
\label{kappat-1}
\ee
In the limit $y\to0$ \eref{dsn2t} is recovered. The mean value in~\eref{xmean} and the variance are 
compared to finite-size data in figure~\ref{fig-4}.

\begin{figure}[!th]
\begin{center}
\includegraphics[width=8cm,angle=0]{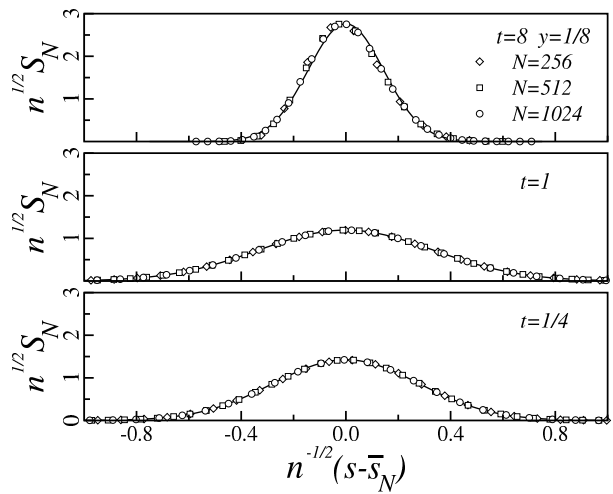}
\end{center}
\vglue -.5cm
\caption{
\label{fig-6} As in figure~\ref{fig-5} for $y=\rho_A-\rho_B=1/8$.
}
\end{figure}

The scaling behaviour of the probability distribution $S_N(s,k)$ at different times 
is shown in figure~\ref{fig-5} for $\rho_A=\rho_B$ and figure~\ref{fig-6} for 
$\rho_A-\rho_B=1/8$.

\section{Time required to reach a given number of surviving particles}
\subsection{Probability distribution}

This section is dedicated to the study of $T_N(s,k)$, the probability distribution for the 
number of updates $k$ needed to reach for the first time a total number 
of surviving particles $s_A+s_B=2s$, 
as shown in figure~\ref{fig-7}(a). This probability is related to $S_N(s+1,k-1)$
through:
\be
T_N(s,k)=S_N(s+1,k-1)\times 2\frac{(s+1)^2-d^2}{N^2}\,.
\label{tnsk-1}
\ee
It is given by the product of the probability to be in a state with $s'=s+1$ 
after $k-1$ updates by the probability of the transition 
$s'=s+1\rightarrow~s''=s$ at the next update.
Making use of~\eref{snsk} one obtains:
\be\fl
T_N(s,k)=\frac{(n-d)!(n+d)!}{n^2(s-d)!(s+d)!}\sum_{r=s+1}^n
\frac{(-1)^{r-s-1}r(r+s)!}{(n-r)!(n+r)!(r-s-1)!}
\left(1\!-\!2\,\frac{r^2\!-\!d^2}{N^2}\right)^{k-1}\!\!\!\!\,.
\label{tnsk-2}
\ee
The evolution with $k$ is governed by the master equation
\be\fl
T_N(s,k)=\left[1-2\,\frac{(s+1)^2-d^2}{N^2}\right]\!T_N(s,k-1)
+2\,\frac{(s+1)^2-d^2}{N^2}\,T_N(s+1,k-1)\,,
\label{mastert}
\ee
which follows from~\eref{master} and~\eref{tnsk-1}.

\subsection{Generating function}

In order to calculate the mean value and the variance of the reaction time $t=k/N$ 
we introduce the generating function
\be
\CT_N(s,z)=\sum_{k=1}^\infty z^kT_N(s,k)\,.
\label{tnsz-1}
\ee
\begin{figure}[!th]
\begin{center}
\includegraphics[width=10cm,angle=0]{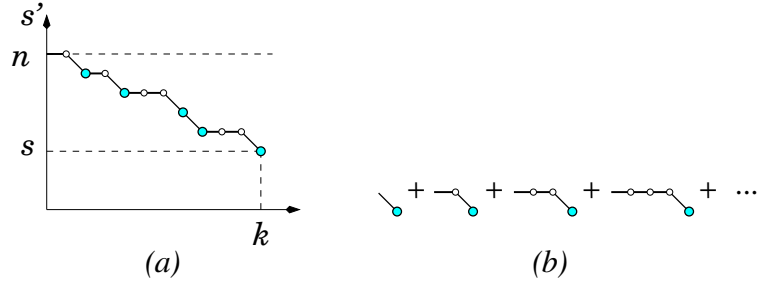}
\end{center}
\vglue -.5cm
\caption{(a) Evolution of $s'=(s'_A+s'_B)/2$ as a function of the number of updates $k$.
Initially all the sites are occupied, $s'=n$. Small circles correspond to the final 
state of updates where $s'$ keeps the same value, bigger circles to a transition $s'\to s'-1$
when two particles annihilate. (b) Diagrams corresponding to the generating function 
$\CL_N(s',z)$ in~\eref{lnsz} for the lifetime of a state with $2s'$ particles. 
\label{fig-7}
}
\end{figure}
The evolution from $s'=n$ to $s'=s$ in figure~\ref{fig-7}(a) proceeds through a succession of steps
where the system remains for some time in a state 
with $s'_A+s'_B=2s'$ until two particles annihilate and $s'\rightarrow~s'-1$.
We associate with such a step the generating function for its lifetime corresponding 
to the diagrams of figure~\ref{fig-7}(b):
\bea
\fl\CL_N(s',z)&=\Bigg\{1+z\left[1-2\frac{s'^2-d^2}{N^2}\right]+\cdots
+\underbrace{z^l\left[1-2\frac{s'^2-d^2}{N^2}\right]^l}_{l\ \mathrm{updates\ without\ annihilation}}
+\cdots\Bigg\}\underbrace{2z\frac{s'^2-d^2}{N^2}}_{\mathrm{annihilation}}\nonumber\\
\fl&=\frac{2z(s'^2-d^2)}{N^2-z\left[N^2-2(s'^2-d^2)\right]}\,.
\label{lnsz}
\eea
The generating function for the reaction time, measured in the number of updates, is obtained 
as the product:
\be\fl
\CT_N(s,z)=\prod_{s'=s+1}^{n}\CL_N(s',z)=\frac{(n-d)!(n+d)!}{(s-d)!(s+d)!}
\,\frac{(2z)^{n-s}}{\prod_{s'=s+1}^n\left\{N^2-z\left[N^2-2(s'^2-d^2)\right]\right\}}\,.
\label{tnsz-2}
\ee
It is easy to verify on this expression that $\CT_N(s,1)=\sum_{k=1}^\infty T_N(s,k)=1$
so that $T_N(s,k)$ is properly normalized.

\subsection{Mean value and variance }
The mean value of the reaction time $t=k/N$ is given by:
\be
\overline{t_N(s)}=\frac{1}{N}\sum_{k=1}^\infty k T_N(s,k)
=\frac{1}{N}\left.\frac{\partial\CT_N}{\partial z}\right|_{z=1}
=n\sum_{s'=s+1}^n\frac{1}{s'^2-d^2}\,.
\label{tns-1}
\ee
When $d=0$ one obtains
\be
\overline{t_N(s)}=n\sum_{s'=s+1}^n\frac{1}{s'^2}=n(H_n^{(2)}-H_s^{(2)})\,,
\label{tns-2}
\ee
where $H_l^{(m)}=\sum_{j=1}^l1/j^m$ is a generalized harmonic number.
When $d>0$ one may write
\be
\overline{t_N(s)}\!=\!\frac{n}{2d}\!\sum_{s'=s+1}^n\!\left(\frac{1}{s'\!-\!d}-\frac{1}{s'\!+\!d}\right)
\!=\frac{n}{2d}(H_{n-d}\!-\!H_{s-d}\!-\!H_{n+d}\!+\!H_{s+d})\,,
\label{tns-3}
\ee
where $H_l=H_l^{(1)}$ is a harmonic number.

A second derivative gives the mean square value of the reaction time:
\bea
\overline{t_N^2(s)}&=\frac{1}{N^2}\sum_{k=1}^\infty k^2 T_N(s,k)
=\frac{1}{N^2}\left.\frac{\partial}{\partial z}
\left(z\frac{\partial\CT_N}{\partial z}\right)\right|_{z=1}\nonumber\\
&=n^2\left(\sum_{s'=s+1}^n\frac{1}{s'^2\!-\!d^2}\!\right)^2
\!\!\!+n^2\!\sum_{s'=s+1}^n\frac{1}{(s'^2\!-\!d^2)^2}-
\frac{1}{2}\!\sum_{s'=s+1}^n\!\frac{1}{s'^2\!-\!d^2}\,.
\label{tn2s-1}
\eea
Since the first term in the last expression is $\overline{t_N(s)}^2$ the variance is given by:
\be
\overline{\Delta t_N^2(s)}=
n^2\!\sum_{s'=v+1}^n\frac{1}{(s'^2\!-\!d^2)^2}-\frac{1}{2}\!\sum_{s'=s+1}^n\!\frac{1}{s'^2\!-\!d^2}\,.
\label{dtn2s-1}
\ee
When $d=0$ one obtains:
\be
\overline{\Delta t_N^2(s)}\!=\!
n^2\!\sum_{s'=s+1}^n\frac{1}{s'^4}\!-\!\frac{1}{2}\!\sum_{s'=s+1}^n\!\frac{1}{s'^2}
\!=\!n^2(H_n^{(4)}\!-\!H_s^{(4)})\!-\!\frac{1}{2}(H_n^{(2)}\!-\!H_s^{(2)})\,.
\label{dtn2s-2}
\ee
When $d>0$ \eref{dtn2s-1} can be rewritten as:
\bea
\fl\overline{\Delta t_N^2(s)}&=\frac{n^2}{4d^2}\sum_{s'=s+1}^n
\left[\frac{1}{(s'-d)^2}+\frac{1}{(s'+d)^2}\right]-\frac{1}{4d}\left(1+\frac{n^2}{d^2}\right)
\sum_{s'=s+1}^n\left(\frac{1}{s'-d}-\frac{1}{s'+d}\right)\nonumber\\
\fl&=\!\frac{n^2}{4d^2}(H_{n\!-\!d}^{(2)}\!-\!H_{s\!-\!d}^{(2)}\!
+\!H_{n\!+\!d}^{(2)}\!-\!H_{s\!+\!d}^{(2)})\!-\!\frac{1}{4d}\!\left(\!1\!+\!\frac{n^2}{d^2}\right)
\!(H_{n\!-\!d}\!-\!H_{s\!-\!d}\!-\!H_{n\!\!+d}\!+\!H_{s\!+\!d})\,.
\label{dtn2s-3}
\eea

\subsection{Scaling limit}
In the scaling limit the probability distribution $T_N(s,k)$ leads to three different 
probability densities,
depending on the values of $x=s/n$ and $y=d/n$. We now study these different cases.

\subsubsection{$x>y\geq 0$.}
\begin{figure}[!th]
\begin{center}
\includegraphics[width=8cm,angle=0]{fig-8.eps}
\end{center}
\vglue -.5cm
\caption{Scaling behaviour of (a) the mean value $\overline{t_N}$ and 
(b) the variance $\overline{\Delta t_N^2}$ of the time $t$ needed 
to reach a given value of $x=\rho_A+\rho_B$ with $\rho_A=\rho_B=1/2$ 
in the initial state. The finite-size data for $N=64$ (diamond), 128 (square), 
256 (circle), given by~\eref{tns-1} and~\eref{dtn2s-1}, collapse on 
the full lines corresponding to the scaling functions in~\eref{tn-1} and~\eref{chi0x}.
\label{fig-8}
}
\end{figure}
\begin{figure}[!th]
\begin{center}
\includegraphics[width=8cm,angle=0]{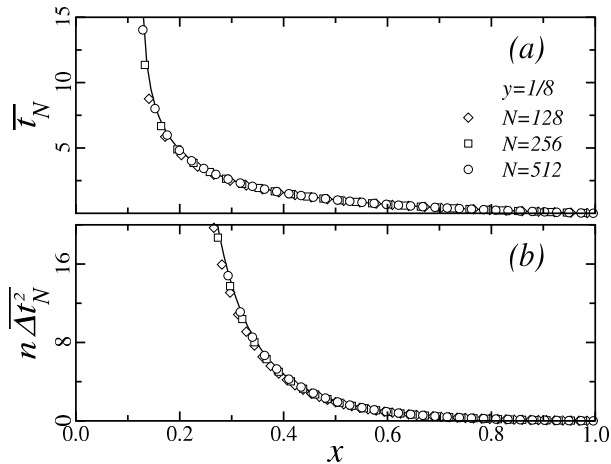}
\end{center}
\vglue -.5cm
\caption{As in figure~\ref{fig-8} for $\rho_A-\rho_B=1/8$ and $N=128$ (diamond),
 256 (square), 512 (circle). The scaling functions are given in~\eref{tns-4}
 and~\eref{dtn2-1}.
\label{fig-9}
}
\end{figure}
We first consider the case where both $n,s,d\to\infty$ for fixed values 
of the ratios $x>y\geq0$. 
Using in~\eref{tns-3} the asymptotic expansion for harmonic numbers
\be
H_{n\alpha}=\ln n+\ln\alpha+\gamma+\Or(n^{-1})\,,
\label{hn}
\ee
where $\gamma$ is Euler's constant, gives
\be
\overline{t_N}\scal h_y(x)=\frac{1}{2y}\ln\left[\frac{(1-y)(x+y)}{(1+y)(x-y)}\right]\,,
\label{tns-4}
\ee
which is the mean-field expression~\eref{tmf-1}. It reduces to
\be
\overline{t_N}\scal\frac{1}{x}-1\,,
\label{tn-1}
\ee
when $y\to0$.

Using the following expansion for generalized harmonic numbers
\be\fl
H_{n\beta}^{(2)}-H_{n\alpha}^{(2)}=\sum_{j=n\alpha+1}^{n\beta}\frac{1}{j^2}
=\int_{n\alpha}^{n\beta}\frac{dj}{j^2}+\Or(n^{-2})
=\frac{1}{n}\left(\frac{1}{\alpha}-\frac{1}{\beta}\right)+\Or(n^{-2})
\label{hn2}
\ee
as well as~\eref{hn}, the scaling limit of the variance follows from~\eref{dtn2s-3} and reads:
\bea
\overline{\Delta t_N^2}&\scal\frac{\chi_y(x)}{n}\nonumber\\
\chi_y(x)&=\frac{1}{2y^2}\!\left(\frac{x}{x^2\!-\!y^2}-\frac{1}{1\!-\!y^2}\right)\!
-\!\frac{1}{4y}\left(1\!+\!\frac{1}{y^2}\right)
\ln\left[\frac{(1\!-\!y)(x\!+\!y)}{(1\!+\!y)(x\!-\!y)}\right]\,.
\label{dtn2-1}
\eea
When $y\to0$ one obtains:
\be
\chi_0(x)=\frac{1}{3x^3}-\frac{1}{2x}+\frac{1}{6}\,.
\label{chi0x}
\ee
Here too the fluctuations are small and $t$ is a self-averaging variable when $x>y$.

The dependence on $x=\rho_A+\rho_B$ of the mean value and the variance of the 
reaction time $t$ is shown in figure~\ref{fig-8} for $y=\rho_A-\rho_B=0$ and 
figure~\ref{fig-9} for $y=1/8$. A good collapse of the finite-size data is obtained. 

\begin{figure}[!th]
\begin{center}
\includegraphics[width=8cm,angle=0]{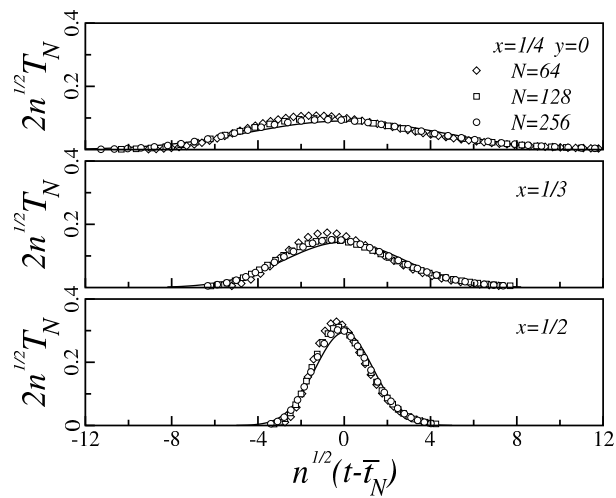}
\end{center}
\vglue -.5cm
\caption{Data collapse for the scaled probability distribution $2n^{1/2}T_N(s,k)$ as 
a function of $\theta=n^{1/2}(t-\overline{t_N(s)})$ at different values of the particle density
$x=\rho_A+\rho_B$ and for increasing lattice sizes, $N=64$ (diamond), 128 (square) and 256 (circle).
In the initial state $\rho_A=\rho_B=1/2$. The full lines correspond to the Gaussian 
density, $\fT(x,\theta)$ in~\eref{txtheta-2}, which is obtained in the scaling limit.
The fluctuations are growing as $x$ decreases.
\label{fig-10}
}
\end{figure}

\begin{figure}[!th]
\begin{center}
\includegraphics[width=8cm,angle=0]{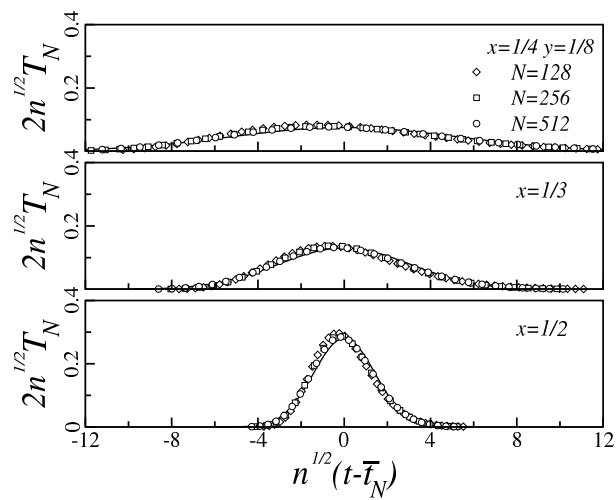}
\end{center}
\vglue -.5cm
\caption{As in figure~\ref{fig-10} for $y=\rho_A-\rho_B=1/8$ with $N=128$ (diamond), 
256 (square), 512 (circle).
\label{fig-11}
}
\end{figure}

The scaling of the variance with $n$ suggests the definition of the 
following scale-invariant and centered time variable
\be
\theta(s,k)=n^{1/2}(t-\overline{t_N})\,,\qquad t=\frac{k}{2n}
\label{theta}
\ee
where $\overline{t_N}$ given by~\eref{tns-4} depends on $s$ through $x=s/n$.
Accordingly we define the probability density as:
\be
\fT(x,\theta)\scal2n^{1/2}T_N(s,k)\,.
\label{txtheta-1}
\ee
We proceed as in section~\ref{sec-scalsn}~\footnote[2]{Except that here the expression
of $\overline{t_N}$ is already known.} and solve the master 
equation~\eref{mastert} in the scaling limit. 
\begin{figure}[!t]
\begin{center}
\includegraphics[width=8cm,angle=0]{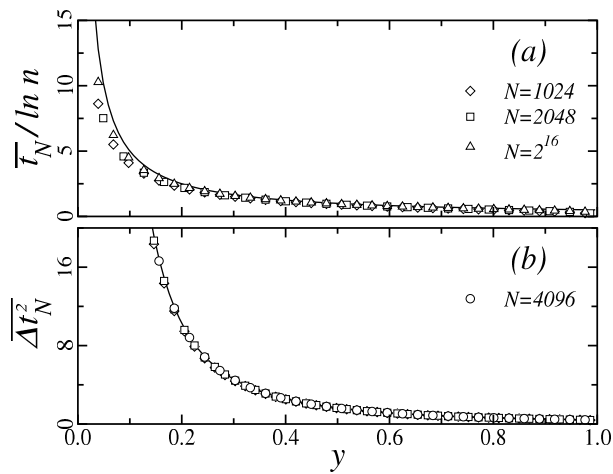}
\end{center}
\vglue -.5cm
\caption{Scaling behaviour of (a) the mean value $\overline{t_N}$ and 
(b) the variance $\overline{\Delta t_N^2}$ of the time $t$ needed 
to reach a state with $y=x=\rho_A$ and $u=\rho_B=0$ with $\rho_A=(1+y)/2$ 
and $\rho_B=(1-y)/2$ in the initial state. The finite-size data for $N=1024$ 
(diamond), 2048 (square), 4096 (circle), $2^{16}$ (triangle), given 
by~\eref{tns-1} and~\eref{dtn2s-1}, collapse on 
the full lines corresponding to the scaling functions in~\eref{tn-3} 
and~\eref{dtn2-2}. The convergence is slow for $\overline{t_N}$ due to a 
correction to scaling of order $(\ln n)^{-1}$.
\label{fig-12}
}
\end{figure}
We replace $T_N$ by $\fT$,  
make a Taylor expansion of $\fT$ to second order in $s$ and $k$ and rewrite 
the coefficients and the derivatives in terms of the new scaled variables.
In this way an expansion in powers of $n^{-1/2}$ is obtained. The coefficients of $n^0$  
and $n^{-1/2}$ vanish identically. The first non-vanishing contribution is coming at order $n^{-1}$ 
and leads to the partial differential equation:
\be
\frac{\partial\fT}{\partial x}=\frac{1}{2(x^2-y^2)}
\left(\frac{1}{2}-\frac{1}{x^2-y^2}\right)\frac{\partial^2\fT}{\partial\theta^2}\,.
\label{dtdx}
\ee
Assuming that $\fT$ depends on $x$ only through the reduced variance $\chi_y(x)$
given by~\eref{dtn2-1} the partial differential equation transforms into the 
diffusion equation
\be
\frac{\partial\fT}{\partial\chi_y}=\frac{1}{2}\frac{\partial^2\fT}{\partial\theta^2}\,.
\label{dtdchi}
\ee
Thus the fluctuations are Gaussian:
\be
\fT(x,\theta)=\frac{\e^{-\theta^2/[2\chi_y(x)]}}{\sqrt{2\pi\chi_y(x)}}\,,\qquad x>y\geq0\,. 
\label{txtheta-2}
\ee

The Gaussian  behaviour of the reaction time is shown for different values of 
$x=\rho_A+\rho_B$ in figure~\ref{fig-10} for $\rho_A=\rho_B$ and figure~\ref{fig-11}
for $\rho_A-\rho_B=1/8$. The finite-size results were obtained by iterating the master 
equation~\eref{master}, storing the data at each update for a given value of $s=nx+1$ 
and using~\eref{tnsk-1}.

\subsubsection{$x=y>0$, $s\geq d$.}
We study now the case where $s=d+u$ with $d=ny$, $s_A=\Or(n)$ and $s_B=u=\Or(1)$, 
i.e. when the reaction is close to completion or complete. 
The mean reaction time which follows from~\eref{tns-3}  
\be 
\overline{t_N}=\frac{1}{2y}\left[\ln\left(2ny\frac{1-y}{1+y}\right)
+\gamma-H_u\right]+\Or(n^{-1})\,,
\label{tn-2}
\ee
has a slow logarithmic growth with $n$ which in the scaling limit yields:
\be
\frac{\overline{t_N}}{\ln n}=\frac{1}{2y}+\Or[(\ln n)]^{-1})\,.
\label{tn-3}
\ee
The variance in~\eref{dtn2s-3} gives
\be
\overline{\Delta t_N^2}=\frac{1}{4y^2}\left[\zeta(2)-H_u^{(2)}\right]
+\Or\left(\frac{\ln n}{n}\right)\,.
\label{dtn2-2}
\ee
The fluctuations of the reaction time $t$ are stronger when the reaction is completed 
or close to completion. The finite-size data collapse for the mean value 
and the variance as a function of $y$ is shown in figure~\ref{fig-12}.

In this regime the ratio $\sqrt{\overline{\Delta t_N^2}}/\overline{t_N}$ decreases slowly 
as $(\ln n)^{-1}$
which suggests a new type of statistics. It will be obtained by taking the scaling limit
directly on the probability distribution $T_N(s,k)$ in~\eref{tnsk-2}.
Since the variance does not depend on $n$ we define a centered time variable as:
\be
\theta'=2y(t-\overline{t_N})\,,\qquad t=\frac{k}{2n}\,.
\label{theta'}
\ee
The factor $2y$ is suggested by the form of $\overline{t_N}$ given by~\eref{tn-2}.
With this definition one obtains:
\be
k=\frac{n}{y}\left[\theta'+\ln\left(2ny\frac{1-y}{1+y}\right)+\gamma-H_u\right]\,.
\label{k}
\ee
Thus one defines the probability density as:
\be
\fT'(u,\theta')\scal\frac{n}{y}T_N(d+u,k)\,.
\label{tutheta'-1}
\ee
With the change of summation variable $r=j+d+u+1$~\eref{tnsk-2} leads to
\be
\frac{n}{y}T_N(d+u,k)=\frac{1}{u!}\sum_{j=0}^{n-d-u-1}\frac{(-1)^j}{j!}
A_jB_jC_jD_j\left(1-E_j\right)^{k-1}\,,
\label{nytn}
\ee
where, in the scaling limit with $d=ny$:
\be
A_j=\frac{d+j+u+1}{d}=1+\frac{j+u+1}{ny}\simeq1\,,
\ee
\be
B_j=\frac{(n\!-\!d)!}{(n\!-\!d\!-\!j\!-\!u\!-\!1)!}
=\prod_{i=0}^{j+u}(n-d-i)\simeq(n\!-\!d)^{j+u+1}\,,
\ee
\be
C_j=\frac{(n\!+\!d)!}{(n\!+\!d\!+\!j\!+\!u\!+\!1)!}
=\left[\prod_{i=1}^{j+u+1}(n+d+i)\right]^{-1}
\simeq(n\!+\!d)^{-(j+u+1)}\,,
\ee
\be
D_j=\frac{(2d\!+\!2u\!+\!j\!+\!1)!}{(2d\!+\!u)!}
=\prod_{i=1}^{j+u+1}(2d+u+i)
\simeq(2d)^{j+u+1}\,,
\ee
\be
E_j=\frac{(d\!+\!j\!+\!u\!+\!1)^2-d^2}{2n^2}
=\frac{2d\!+\!j\!+\!u\!+\!1}{2n^2}\,(j\!+\!u\!+\!1)
\simeq\frac{y}{n}\,(j\!+\!u\!+\!1)\,.
\ee
\begin{figure}[!t]
\begin{center}
\includegraphics[width=8cm,angle=0]{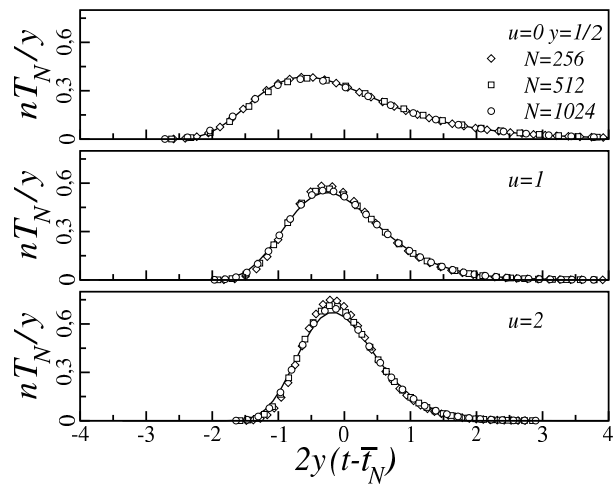}
\end{center}
\vglue -.5cm
\caption{Data collapse for the scaled probability distribution $nT_N(s,k)/y$ as 
a function of $\theta'=2y(t-\overline{t_N(s)})$ at different values of $u=s_B$,
for $y=\rho_A-\rho_B=1/2$ and increasing lattice sizes, $N=256$ (diamond), 
512 (square) and 1024 (circle). The full lines correspond to the 
generalized Gumbel distribution in~\eref{tutheta'-3}, which crosses over to a Gaussian
as $u$ increases.
\label{fig-13}
}
\end{figure}
Taking into account the expression of $k$ in~\eref{k} one obtains:
\be
\left(1-E_j\right)^{k-1}\simeq\left(2ny\frac{1-y}{1+y}\right)^{-(j+u+1)}
\e^{-(j+u+1)(\theta'+\gamma-H_u)}
\ee
Thus in the scaling limit~\eref{nytn} gives
\be
\fT'(u,\theta')=\frac{1}{u!}\e^{-(u+1)(\theta'+\gamma-H_u}\sum_{j=0}^\infty
\frac{\left[-\e^{-(\theta'+\gamma-H_u}\right]^j}{j!}\,,
\label{tutheta'-2}
\ee
which is the generalized Gumbel distribution~\cite{ojo2001,pinheiro2016}:
\be
\fT'(u,\theta')=\frac{1}{u!}\exp\left[-(u+1)(\theta'+\gamma-H_u)
-\e^{-(\theta'+\gamma-H_u)}\right]\,.
\label{tutheta'-3}
\ee
It crosses over to the Gaussian in~\eref{txtheta-2} when $u\gg1$. 
The matching of the two probability densities is studied in appendix E.
The collapse of the finite-size data on the generalized Gumbel distribution is shown 
in figure~\ref{fig-13} for different values of $u$ and $y=\rho_A-\rho_B=1/2$.
The finite-size data were obtained by iterating~\eref{master} and using~\eref{tnsk-1}.

\subsubsection{$x=y=0$, $s\geq d$.}
\begin{figure}[!th]
\begin{center}
\includegraphics[width=8cm,angle=0]{fig-14.eps}
\end{center}
\vglue -.5cm
\caption{Scaling behaviour of (a) the mean value $\overline{t_N}$ and 
(b) the variance $\overline{\Delta t_N^2}$ of the time $t$ needed 
to reach a given value of $s=(s_A+s_B)/2$ with $s_A=s_B=n$ 
in the initial state. The finite-size data for $N=64$ 
(diamond), 128 (square), 256 (circle), given by~\eref{tns-1} and~\eref{dtn2s-1}, 
given by~\eref{tns-1} and~\eref{dtn2s-1}, collapse on the full lines 
corresponding to the scaling functions in~\eref{tn-dtn2-1}.
\label{fig-14}
}
\end{figure}
\begin{figure}[!th]
\begin{center}
\includegraphics[width=8cm,angle=0]{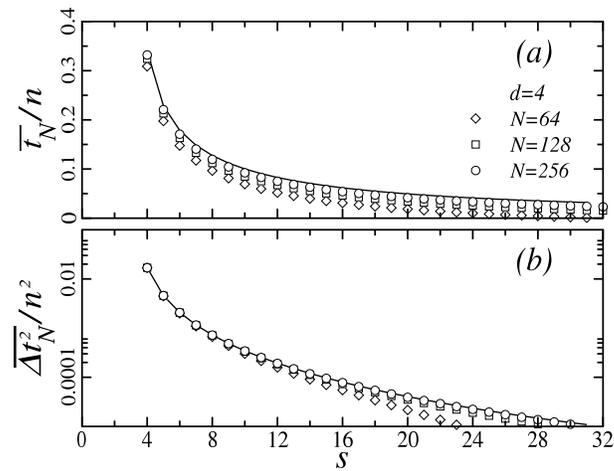}
\end{center}
\vglue -.5cm
\caption{As in figure~\ref{fig-14} with $s_A=n+4$ and $s_B=n-4$ in the initial state. 
The scaling functions are now given by~\eref{tn-dtn2-2}.
\label{fig-15}
}
\end{figure}
Finally we consider the case when $s=\Or(1)\geq d$. According to~\eref{tns-2} 
and~\eref{dtn2s-2} in the scaling limit the mean value and the variance of the 
reaction time behave as 
\be
\frac{\overline{t_N}}{n}\scal\zeta(2)-H_s^{(2)}\,,\qquad 
\frac{\overline{\Delta t_N^2}}{n^2}\scal\zeta(4)-H_s^{(4)}\,,
\label{tn-dtn2-1}
\ee
when $d=0$ whereas~\eref{tns-3} and~\eref{dtn2s-3} lead to
\be\fl
\frac{\overline{t_N}}{n}\scal\frac{1}{2d}(H_{s+d}-H_{s-d})\,,\quad 
\frac{\overline{\Delta t_N^2}}{n^2}\scal\frac{1}{4d^2}
\left[2\zeta(2)\!-\!H_{s-d}^{(2)}\!-\!H_{s+d}^{(2)}\!+\!\frac{1}{d}(H_{s-d}\!-\!H_{s+d})\right],
\label{tn-dtn2-2}
\ee
when $d>0$. The finite-size data collapse is shown in figure~\ref{fig-14} for $s_A=s_B$ 
and figure~\ref{fig-15} for $d=(s_A-s_B)/2=4$.

The mean value and the standard deviation are both growing as $n$, 
thus the reaction time is a strongly fluctuating random variable 
when the reaction is almost complete. In the following we use the 
scale-invariant time variable
\be
\theta''=\frac{t}{n}=\frac{k}{2n^2}\,,
\label{theta''}
\ee
and define the associated probability density as:
\be
\fT''(s,\theta'')\scal2n^2T_N(s,k)\,.
\label{tstheta''-1}
\ee
\begin{figure}[!th]
\begin{center}
\includegraphics[width=8cm,angle=0]{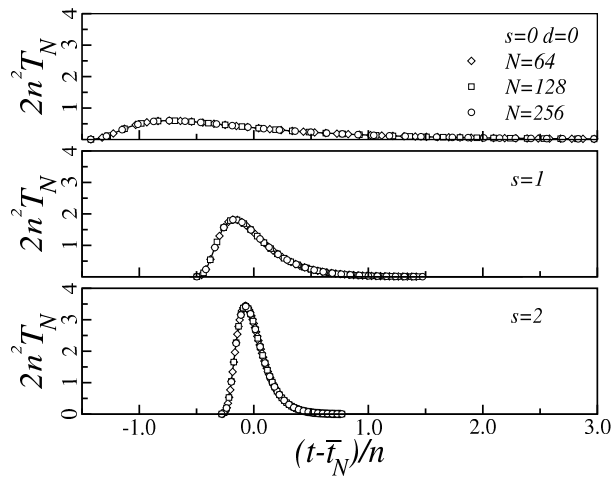}
\end{center}
\vglue -.5cm
\caption{Data collapse for the scaled probability distribution $2n^2T_N(s,k)$ as 
a function of $\theta''-\overline{\theta''}=(t-\overline{t_N(s)})/n$ 
at different values of $s=(s_A+s_B)/2$,
for $s_A=s_B=n$ in the initial state and increasing lattice sizes, $N=64$ (diamond), 
128 (square) and 256 (circle). The full lines correspond to the 
probability density $\fT''(s,\theta''-\overline{\theta''})$ in~\eref{tstheta''-2}, 
which crosses over to a Gaussian
as $s$ increases.
\label{fig-16}
}
\end{figure}
\begin{figure}[!th]
\begin{center}
\includegraphics[width=8cm,angle=0]{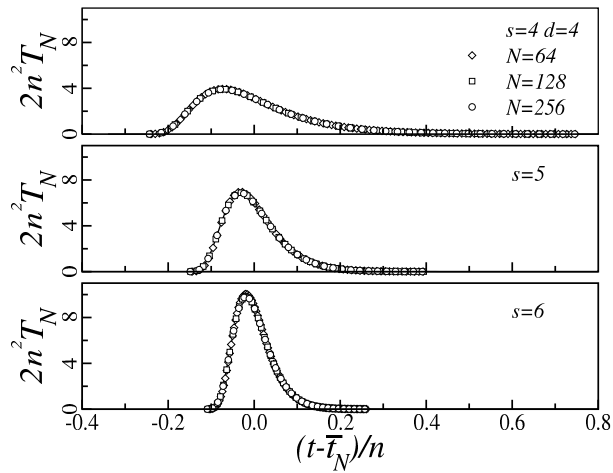}
\end{center}
\vglue -.5cm
\caption{As in figure~\ref{fig-16} with $s_A=n+4$ and $s_B=n-4$ in the initial state.
\label{fig-17}
}
\end{figure}
In the expression~\eref{tnsk-2} of $T_N(s,k)$ one may write:
\be
\frac{r(r+s)!}{(r-s-1)!}=(-1)^{s+1}\prod_{m=0}^s(m^2-r^2)\,.
\label{prodm}
\ee
Furthermore, in the scaling limit, one obtains:
\be\fl
\left(1-2\frac{r^2-d^2}{N^2}\right)^{k-1}\simeq\e^{-(r^2-d^2)\theta''}\,,\qquad
\frac{(n-d)!(n+d)!}{(n-r)!(n+r)!}=\prod_{j=d}^{r-1}\frac{n-j}{n+j+1}\simeq1\,.
\label{limits}
\ee
Thus the probability density is given by: 
\be
\fT''(s,\theta'')=\frac{2\,\e^{d^2\theta''}}{(s-d)!(s+d)!}
\sum_{r=s+1}^\infty(-1)^r\prod_{m=0}^s(m^2-r^2)\,\e^{-r^2\theta''}\,.
\label{tstheta''-2}
\ee
\begin{figure}[!th]
\begin{center}
\psfrag{Y}[Bc][Bc][1][1]{$\mathfrak{T}''(s,\theta'')$}
\psfrag{X}[tc][tc][1][0]{$\theta''$}
\psfrag{a}[Bc][Bc][1][1]{\tiny $s=0$}
\psfrag{b}[Bc][Bc][1][1]{\tiny $s=1$}   
\psfrag{c}[Bc][Bc][1][1]{\tiny $s=2$}   
\psfrag{d}[Bc][Bc][1][1]{\tiny $d=0$}   
\includegraphics[width=8cm,angle=0]{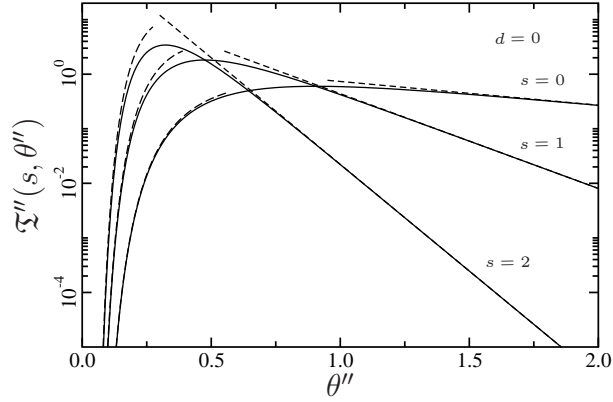}
\end{center}
\vglue -.5cm
\caption{Semi-logarithmic plot of $\fT(s,\theta'')$ in~\eref{tstheta''-2} for $s=0,1,2$ 
with $s_A=s_B=n$ in the initial state. The dashed lines correspond to the 
asymptotic behaviour in~\eref{tstheta''-3} for $\theta''\ll1$ and~\eref{tstheta''-6} 
for $\theta''\gg1$. A similar agreement is obtained when $d>0$.
\label{fig-18}
}
\end{figure}
The collapse of the finite-size data on $\fT''(s,\theta'')$ is shown at different 
values of $s$ for $s_A=s_B$ in figure~\ref{fig-16} and for $d=(s_A-s_B)/2=4$ in figure~\ref{fig-17}.
The finite-size data were obtained by iterating~\eref{master} and using~\eref{tnsk-1}.

The asymptotic behaviour for $\theta''\gg1$ (see figure~\ref{fig-18}) is governed 
by the first term in the sum and reads:
\be 
\fT''(s,\theta'')\simeq2(s+1)(2s+1){2s\choose s+d}\e^{-[(s+1)^2-d^2]\theta''}\,,
\qquad\theta''\gg1\,.
\label{tstheta''-3}
\ee

In order to study the asymptotic behaviour when $\theta''\ll1$ it will be convenient to 
re-express $\fT''(s,\theta'')$ in terms of Jacobi theta functions. 
First let us notice that the product in~\eref{tstheta''-2} vanishes for $r\leq s$ 
so that the sum can start at $r=1$ instead of $s+1$. The product can then be replaced 
by the differential operator $\prod_{m=0}^s(m^2+d/d\theta'')$ so that: 
\be
\sum_{r=s+1}^\infty(-1)^r\prod_{m=0}^s(m^2-r^2)\,\e^{-r^2\theta''}
=\prod_{m=0}^s(m^2+d/d\theta'')\sum_{r=1}^\infty(-1)^r\e^{-r^2\theta''}\,.
\label{sumr-1}
\ee
Making use of the identity~(\cite{whittaker27} p 463)
\be
\sum_{r=1}^\infty(-1)^rq^{r^2}\cos(2rz)=\frac{\vartheta_4(z,q)-1}{2}\,,
\label{sumr-2}
\ee
one obtains
\be
\fT''(s,\theta'')=\frac{\e^{d^2\theta''}}{(s-d)!(s+d)!}
\prod_{m=0}^s(m^2+d/d\theta'')\,\vartheta_4\left(0,\e^{-\theta''}\right)\,,
\label{tstheta''-4}
\ee
where the Jacobi theta function can be re-written as~(\cite{whittaker27} p 475):
\be
\vartheta_4\left(0,\e^{-\theta''}\right)
=\sqrt{\frac{\pi}{\theta''}}\vartheta_2\left(0,\e^{-\pi^2/\theta''}\right)
=2\sqrt{\frac{\pi}{\theta''}}\sum_{r=0}^\infty\e^{-(r+1/2)^2\pi^2/\theta''}\,.
\label{theta4thetha2}
\ee
Finally the probability density takes the following form:
\be
\fT''(s,\theta'')=\frac{2\e^{d^2\theta''}}{(s\!-\!d)!(s\!+\!d)!}
\prod_{m=0}^s(m^2\!+\!d/d\theta'')\sqrt{\frac{\pi}{\theta''}}
\sum_{r=0}^\infty\e^{-(r+1/2)^2\pi^2/\theta''}\!\!.
\label{tstheta''-5}
\ee

The leading contribution when $\theta''\ll1$ (see figure~\ref{fig-18}) 
comes from the derivatives of the exponential in the first term of the sum so that:
\be
\fT''(s,\theta'')\simeq\frac{2\sqrt{2}}{(s\!-\!d)!(s\!+\!d)!}
\left(\frac{\pi}{2\theta''}\right)^{2s+5/2}\e^{-\pi^2/(4\theta'')}\,,\qquad \theta''\ll1\,.
\label{tstheta''-6}
\ee

\section{Conclusion}

The two-species diffusion-annihilation process $A+B\rightarrow$~\O\  has been 
studied on the fully-connected lattice with size $N$ for either equal or 
different numbers of particles, $s_A$ and $s_B$, in the initial state where $s_A+s_B=N=2n$.
Exact probability distributions, $S_N$ for $s=(s_A+s_B)/2$ at a given time and 
$T_N$ for the reaction time $t$ needed to reach a given number of surviving particles, have been 
obtained by solving the master equation. The finite-size scaling behaviour of the 
mean value and the variance of $s$ and $t$ has been determined using a generating 
function approach. 

In the scaling limit, the fluctuations of the number of particles around their mean values, 
given exactly by mean-field theory, are weak and Gaussian.
The statistical properties of the reaction time display 
three different regimes. When both $s_A$ and $s_B$ are $\Or(n)$ the fluctuations 
of the reaction time, as measured by the ratio $R=\sqrt{\overline{\Delta t_N^2}}/\overline{t_N}$,
decay as $n^{-1/2}$. The fluctuations are Gaussian and mean-field theory is exact. 
For unequal initial densities, in the vicinity of the absorbing state when
$s_A=\Or(n)$ and $s_B=\Or(1)$, one obtains $R=\Or[(\ln n)^{-1}]$ thus the fluctuations 
are ``marginally weak''. They are governed by a generalized Gumbel distribution, 
indexed by $s_B$, which crosses over to the Gaussian density with increasing values of $s_B$.
For equal or almost equal initial numbers of particles and when the system is
close to the absorbing state, i.e. when $s_A=\Or(1)\geq s_B\geq0$, one obtains $R=\Or(1)$.
The reaction time is then strongly fluctuating. Its probability density involves 
an alternating infinite series which can be considered as resulting from the applications 
of a product of $s+1$ first-order differential operators to a Jacobi theta function.

A generalized Gumbel distribution has been recently shown to govern the fluctuations of the 
covering time of the fully-connected lattice, i.e. the time needed for a random 
walker to visit almost each site at least once~\cite{turban15} (see also~\cite{chupeau15}). 
Although the two problems present some analogies (they share the same value of $R$ 
and the same probability density), the mean values and the variances scale differently 
with $N$, the number of random walkers is constant for the covering time and 
decreasing for the reaction time.

The extreme value statistics obtained for the reaction time near the absorbing state, 
for almost equal initial numbers of $A$ and $B$ particles, is quite similar 
to what was obtained in~\cite{turban18} for the coagulation process $A+A\rightarrow\ A$. 
Probability densities with nearly the 
same form of asymptotics govern the behaviour of the squared width of an interface generated 
by a periodic Brownian motion~\cite{foltin94} as well as the maximum height of the 1D Edwards-Wilkinson 
model~\cite{edwards82,majumdar04,majumdar05a,majumdar05b,fortin15}.

It seems reasonable to conjecture that similar extreme value statistics should govern 
the reaction time when the system is close to its absorbing state for $D>D_{\mathrm c}=2$
when $s_A-s_B=\Or(n)$ in the initial state and for $D>D_{\mathrm{seg}}=4$ when $s_A-s_B=\Or(1)$
in the initial state.

\appendix

\section{Evaluation of \boldmath{$v_r^{(r)}$}}
According to~\eref{sns0}, in the initial state:
\be
S_N(n,0)=1=v_n^{(n)}\,.
\label{vnn}
\ee
The recursion relation~\eref{rec} gives
\be
S_N(n-1,0)=0=v_{n-1}^{(n-1)}+v_{n-1}^{(n)}=v_{n-1}^{(n-1)}-\frac{n^2-d^2}{n^2-(n-1)^2}\,,
\label{snn-1}
\ee
so that:
\be
v_{n-1}^{(n-1)}=\frac{n^2-d^2}{n^2-(n-1)^2}\,.
\label{vn-1n-1}
\ee
In the same way
\be
S_N(n-2,0)=0=v_{n-2}^{(n-2)}+v_{n-2}^{(n-1)}+v_{n-2}^{(n)}
\label{snn-2}
\ee
where
\bea
\fl v_{n-2}^{(n-1)}&=-\frac{(n-1)^2-d^2}{(n-1)^2-(n-2)^2}v_{n-1}^{(n-1)}
=-\frac{(n^2-d^2)[(n-1)^2-d^2]}{[n^2-(n-1)^2][(n-1)^2-(n-2)^2]}\,,\nonumber\\
\fl v_{n-2}^{(n)}&=-\frac{(n-1)^2-d^2}{n^2-(n-2)^2}v_{n-1}^{(n)}
=\frac{(n^2-d^2)[(n-1)^2-d^2]}{[n^2-(n-1)^2][n^2-(n-2)^2]}\,,
\label{vn-2n-1}
\eea
so that:
\be
v_{n-2}^{(n-2)}=\frac{(n^2-d^2)[(n-1)^2-d^2]}{[n^2-(n-2)^2][(n-1)^2-(n-2)^2]}\,.
\label{vn-2n-2}
\ee
After lengthy but straitforward calculations the same procedure leads to:
\be
v_{n-3}^{(n-3)}=\frac{(n^2-d^2)[(n-1)^2-d^2][(n-2)^2-d^2]}{[n^2-(n-3)^2]
[(n-1)^2-(n-3)^2][(n-2)^2-(n-3)^2]}\,.
\label{vn-3n-3}
\ee
These results suggest the following conjecture
\be
v_r^{(r)}=\prod_{j=0}^{n-r-1}\frac{(n-j)^2-d^2}{(n-j)^2-r^2}\,,
\qquad r=d,\ldots,n-1\,,\qquad v_n^{(n)}=1\,, 
\label{vrr-1}
\ee
from which~\eref{vrr} can be deduced.

\section{Calculation of \boldmath{$\Omega_{r,0}(w)$} and its first derivative at \boldmath{$w=1$}}
According to~\eref{ordw} when $d=0$ 
\be
\Omega_{r,0}(w)=\sum_{s=0}^r(-1)^{r-s}{r+s\choose s}{r\choose s}\frac{2sw^s}{r+s}\,.
\label{or0w-1}
\ee
Using the following combinatorial identity for 
Legendre polynomials~\cite{riordan79}
\be
P_r(2w-1)=\sum_{s=0}^r(-1)^{r-s}{r+s\choose s}{r\choose s}w^s
\label{prw}
\ee
\eref{or0w-1} can be rewritten as:
\be
\Omega_{r,0}(w)=2w^{-r}\!\int_0^w\frac{dP_r(2u-1)}{du}u^rdu\,.
\label{or0w-2}
\ee
An integration by parts gives:
\be
\Omega_{r,0}(w)=2P_r(2w-1)-2rw^{-r}\int_0^wP_r(2u-1)u^{r-1}du\,.
\label{or0w-3}
\ee

When $w=1$ one has:
\be
\Omega_{r,0}(1)=2-2r\!\int_0^1P_r(2u-1)\,u^{r-1}du
=2-r\!\int_{-1}^1\left(\frac{x+1}{2}\right)^{r-1}\!\!\!\!P_r(x)dx\,.
\label{or01-2}
\ee
According to the identity~\cite{gradshteyn80}
\be
\int_{-1}^1(1+x)^\sigma P_\nu(x)dx=
\frac{2^{\sigma+1}\left[\Gamma(\sigma+1)\right]^2}
{\Gamma(\sigma+\nu+2)\Gamma(\sigma-\nu+1)}\,,
\label{intpnu}
\ee
the integral in~\eref{or01-2} vanishes and \eref{or01-1} is obtained.

The first derivative of \eref{or0w-3} at $w=1$ gives:
\be
\left.\frac{d\Omega_{r,0}}{dw}\right|_{w=1}
=2\left.\frac{dP_r(2w-1)}{dw}\right|_{w=1}-2r
\label{dor01-2}
\ee
The generating function for Legendre polynomials
\be
\sum_{l=0}^\infty P_l(x)u^l=\frac{1}{\sqrt{1-2ux+u^2}}
\label{gfpl}
\ee
leads to:
\be
\sum_{l=0}^\infty\left.\frac{dP_l(x)}{dx}\right|_{x=1}u^l=\frac{u}{(1-u)^3}
=\sum_{j=0}^\infty{j+2\choose 2}u^{j+1}\,.
\label{gfdpl}
\ee
Identifying the coefficients of $u^r$, one obtains
\be
\left.\frac{dP_r(2w-1)}{dw}\right|_{w=1}\!\!\!=2\,{r+1\choose 2}=r(r+1)\,,
\label{dpr1}
\ee
and~\eref{dor01-2} leads to~\eref{dor01-1}.

\section{Scaling limit of \boldmath{$\overline{s_N}/n$} and 
\boldmath{$\overline{s_N^2}/n^2$} when \boldmath{$d=0$}}

According to~\eref{snk} one may write: 
\be
\frac{\overline{s_N(k)}}{n}=\sum_{r=1}^n f(r)\,,\qquad f(r)
=\frac{2r}{n-r}\underbrace{\prod_{j=1}^r\frac{n-j}{n+j}}_{p(r)}\,\,
\underbrace{\left(\!1\!-\!2\,\frac{r^2}{N^2}\right)^k}_{e_k(r)}\,.
\label{snkn}
\ee
Making use of the following expansions in powers of $n^{-1}$
\be
\frac{2r}{n-r}\simeq\frac{2r}{n}\left(1+\frac{r}{n}+\frac{r^2}{n^2}\right)\,,
\label{2rnmr}
\ee
\bea
\ln p(r)&=\sum_{j=1}^r\left[\ln\left(1-\frac{j}{n}\right)-\ln\left(1+\frac{j}{n}\right)\right]
\simeq-2\sum_{j=1}^r\left(\frac{j}{n}+\frac{j^3}{3n^3}\right)\nonumber\\
&\simeq-\frac{r(r+1)}{n}-\frac{r^2(r+1)^2}{6n^3}\,,
\label{lnpr}
\eea
\be
\ln e_k(r)=k\ln\left(1-2\frac{r^2}{N^2}\right)
\simeq-2k\left(\frac{r^2}{N^2}+\frac{r^4}{N^4}\right)
\simeq-t\left(\frac{r^2}{n}+\frac{r^4}{4n^3}\right),
\label{lnekr}
\ee
one obtains:
\be
f(r)\!=\!\frac{2r}{n}\left(1\!+\!\frac{r}{n}\!+\!\frac{r^2}{n^2}\right)
\exp\left[-(t\!+\!1)\frac{r^2}{n}\!-\!\frac{r}{n}\!-\!\frac{r^2(r\!+\!1)^2}{6n^3}
-\frac{tr^4}{4n^3}\right]\,.
\label{fr}
\ee
In the scaling limit the Euler-Maclaurin summation formula gives
\be
\sum_{r=1}^n f(r)\simeq\!\int_0^\infty\!\! f(r)\,dr\!+\!\frac{1}{2}\,[f(\infty)-f(0)]
\!+\!\frac{1}{12}\,[f'(\infty)-f'(0)]+\ldots\,,
\label{eml}
\ee
with
\be
f(0)=f(\infty)=f'(\infty)=0\,,\qquad f'(0)=\frac{2}{n}\,.
\label{f0}
\ee
With the change of variable $u=r^2/n$ \eref{fr} and \eref{eml} lead to:
\bea
\fl\sum_{r=1}^n f(r)&\simeq\int_0^\infty\!\!du\left(1\!+\!\sqrt{\frac{u}{n}}\!+\!\frac{u}{n}\right)
\left(1\!-\!\sqrt{\frac{u}{n}}\!-\!\frac{u^2}{6n}\!
-\!\frac{tu^2}{4n}+\!\frac{u}{2n}\right)\e^{-(t+1)u}-\frac{1}{6n}\nonumber\\
\fl&\simeq\int_0^\infty\!\!du\left[1\!+\!\frac{u}{2n}-
\frac{(3t+2)u^2}{12n}\right]\e^{-(t+1)u}-\frac{1}{6n}\,.
\label{sfr}
\eea
Using $\int_0^\infty\!\!du\,u^a\e^{-bu}=a!/b^{a+1}$ one obtains 
$\overline{s_N(k)}/n$ as given in~\eref{snt}.

The expression of $\overline{s_N^2(k)}/n^2$ differs from~\eref{snkn} only 
through a factor $u=r^2/n$ in the sum over $r$. Furthermore the correction of order $n^{-1}$
to the integral in~\eref{eml} now vanishes. Thus we can directly modify the integral in~\eref{sfr}
to write:
\be
\frac{\overline{s_N^2(k)}}{n^2}\simeq\int_0^\infty\!\!du\left[u\!+\!\frac{u^2}{2n}
-\frac{(3t+2)u^3}{12n}\right]\e^{-(t+1)u}\,.
\label{sn2kscal}
\ee
Finally, the integration leads to the expression given in~\eref{snt}.

\section{Solution of equation~\eref{dkapdt}}

With $\kappa_y(t)=\alpha_y(t)\beta_y(t)$ \eref{dkapdt} leads to 
\be
\beta_y(\alpha_y'+4g_y\alpha_y)+\beta_y'\alpha_y+f_y=0\,,\qquad 
f_y(t)=\frac{dg_y}{dt}+\frac{1}{2}\left(\frac{dg_y}{dt}\right)^2\,,
\label{abf}
\ee
which transforms into a system of two first-order differential equations:
\be
\left\{
\begin{array}{ll}
 \alpha_y'+4g_y\alpha_y&=0\,,\\
 \ms
 \beta_y'\alpha_y+f_y&=0\,.
\end{array}
\right.
\label{ab}
\ee
Since $g_y(t)$ in~\eref{xmean} can be written as
\be
g_y(t)=y+\frac{G'}{G}\,,\qquad G=1+y-(1-y)\,\e^{-2yt}\,,
\label{gt}
\ee
the first equation in~\eref{ab} gives:
\be
\alpha_y(t)=C_1\,\e^{-4\int^t\!g_y\,dt'}=C_1\,\frac{\e^{-4yt}}{G^4}\,.
\label{a}
\ee
For the second equation one has
\be
\beta_y(t)=-\int^t\!\frac{f_y}{\alpha_y}\,dt'+C_2
\label{b-1}
\ee
Where according to~\eref{abf} and~\eref{gt}
\be
f_y(t)=-\frac{4y^2(1-y^2)\,\e^{-2yt}}{G^2}+\frac{8y^4(1-y^2)^2\,\e^{-4yt}}{G^4}\,.
\label{ft}
\ee
The integration in~\eref{b-1} is straightforward and gives:
\be
\beta_y(t)=\frac{2y(1-y^2)}{C_1}\left[(1+y)^2\,\e^{2yt}\!-(1-y)^2\,\e^{-2yt}\!-4y(1-y^4)t\right]+C_2\,.
\label{b-2}
\ee
The product of~\eref{a} and~\eref{b-2} leads to:
\be\fl
\kappa_y(t)=\frac{2y(1-y^2)\,\e^{-4yt}}{G^4}
\left[(1+y)^2\,\e^{2yt}-(1-y)^2\,\e^{-2yt}-4y(1-y^4)t\right]+C\,\frac{\e^{-4yt}}{G^4}\,.
\label{kappat-2}
\ee
The initial condition $\kappa_y(0)=0$ is satisfied when $C=C_1C_2=-8y^2(1-y^2)$ 
and~\eref{kappat-1} is finally obtained.

\section{Matching of \boldmath{$\fT'(u,\theta')$} when \boldmath{$u\gg1$} 
with \boldmath{$\fT(x,\theta)$} when \boldmath{$x\ll1$}.}

Let us rewrite $\fT'(u,\theta')$ in~\eref{tutheta'-3} as:
\be\fl
\fT'(u,\theta')=\e^{F_a(b)}\,,\quad F_a(b)=-ab-\e^{-b}-\ln\Gamma(a)\,,\quad a=u+1\,,
\quad b=\theta'+\gamma-H_u\,.
\label{fab-1}
\ee
$F_a(b)$ has a maximum at $b_0=-\ln a$ where 
\be 
F_a(b_0)=a\ln a-a-\ln\Gamma(a)\,,\qquad
F_a''(b_0)=-a\,. 
\label{fab-2}
\ee
Since $a=u+1\gg1$ one has
\bea
\ln\Gamma(a)&\simeq a\ln a-a+\frac{1}{2}\ln\left(\frac{2\pi}{a}\right)\,,\qquad
H_u\simeq\ln u+\gamma\,,\nonumber\\
b-b_0&=b+\ln a=\theta'+\gamma-H_u+\ln(u+1)\simeq\theta'\,,
\label{lngamma}
\eea
and the expansion of $F_a(b)$ around $b_0$ gives:
\be
F_a(b)\simeq-\frac{1}{2}\ln\left(\frac{2\pi}{u}\right)-\frac{u}{2}\theta'^2\,.
\label{fab-3}
\ee
Thus when $u\gg1$ the maximum is amplified, which justifies the approximation, and one obtains:
\be
\fT'(u,\theta')\simeq\frac{\e^{-u\theta'^2/2}}{\sqrt{2\pi/u}}\,,\qquad u\gg1\,.
\label{tutheta'-4}
\ee
According to~\eref{theta} and~\eref{theta'} one has
\be
\theta'=\frac{2y}{n^{1/2}}\theta\,,
\label{change}
\ee
and the change of variables yields
\be
\fT(x,\theta)=\frac{2y}{n^{1/2}}\fT'(u,\theta')=
\frac{\e^{-(4uy^2/n)\theta^2/2}}{\sqrt{2\pi/(4uy^2/n)}}\,,\qquad x=y+\frac{u}{n}\,,
\label{txtheta-3}
\ee
i.e. a Gaussian with variance $\overline{\Delta\theta^2}=n/(4uy^2)$.
This expression has to be compared to~\eref{txtheta-2} when $x-y=u/n\ll1$. It is easy to verify
that, in this limit, the variance $\overline{\Delta\theta^2}=\chi_y(x)$ in~\eref{dtn2-1} 
is governed by the first term on the right and reads
\be
\chi_y(x)\simeq\frac{n}{4uy^2}\,,
\label{chiyx}
\ee
as expected.

\section{Matching between Jacobian and Gaussian regimes}
We were not able to put in evidence this matching at the level of the probability densities. 
Thus we shall compare the mean values and the variances in the appropriate limits: 
$x\ll1$ with $x>y\geq0$ for Gauss and $s\gg1$ with $s\gg d\geq0$ for Jacobi.

For the Gaussian density~\eref{x>y-1} and~\eref{x>y-2} yields
\be
\overline{t_N}\simeq\frac{1}{x}\,,\quad y=0\,;\qquad
\overline{t_N}\simeq\frac{1}{2y}\ln\left(\frac{x+y}{x-y}\right)\,,\quad y>0\,.
\label{x>y-4}
\ee
for the mean value and 
\be\fl
\overline{\Delta t_N^2}\simeq\frac{1}{3nx^3}\,,\quad y=0\,;\quad
\overline{\Delta t_N^2}\simeq\frac{1}{4ny^2}\!
\left[\frac{1}{x-y}+\frac{1}{x+y}+\frac{1}{y}
\ln\left(\frac{x\!-\!y}{x\!+\!y}\right)\right]\,,\quad y>0\,,
\label{x>y-5}
\ee
for the variance.

In the Jacobian regime using the expansions 
$H_s^{(l)}\simeq\zeta(l)-s^{1-l}/(l-1)$ for $l>1$ and 
$H_{s\pm d}\simeq\ln(s\pm d)$ in~\eref{x=y=0-1} gives
\be
\overline{t_N}\simeq\frac{n}{s}\,,\quad d=0\,;\qquad
\overline{t_N}\simeq\frac{n}{2d}\ln\left(\frac{s+d}{s-d}\right)\,,\quad d>0\,,
\label{x=y=0-4}
\ee
for the mean value, in agreement with~\eref{x>y-4}. 
Finally, given the above expansions, \eref{x=y=0-2} leads to
\be\fl
\overline{\Delta t_N^2}\simeq\frac{n^2}{3s^3}\,,\quad d=0\,;\qquad
\overline{\Delta t_N^2}\simeq\frac{n^2}{4d^2}
\left[\frac{1}{s-d}+\frac{1}{s+d}+\frac{1}{d}\ln\left(\frac{s-d}{s+d}\right)\right]
\,,\quad d>0\,,
\label{x=y=0-5}
\ee
for the variance, in agreement with~\eref{x>y-5}.

\section*{References}

\begin{thebibliography}{99}

\bibitem{alcaraz94} Alcaraz F, Droz M, Henkel M and Rittenberg V 1994 {\it Ann. Phys.} {\bf 230}
  250

\bibitem{hinrichsen00} Hinrichsen H 2000 {\it Adv. Phys.} {\bf 49} 815

\bibitem{benavraham00} ben-Avraham D and Havlin S 2000 {\it Diffusion and Reactions 
in Fractals and Disordered Systems} (Cambridge: Cambridge University Press)
\bibitem{schutz01} Sch{\"u}tz G 2001 Exactly solvable models for many-body systems far from
  equilibrium {\it Phase Transitions and Critical Phenomena\/} vol~19 ed Domb C
  and Lebowitz J (London: Academic Press) p~1

\bibitem{odor04} \'Odor G 2004 {\it Rev. Mod. Phys.} {\bf 76} 663

\bibitem{henkel08} Henkel M, Hinrichsen H and L{\"u}beck S 2008 {\it Non-equilibrium phase
  transitions: absorbing phase transitions} vol~1 (Heidelberg: Springer)

\bibitem{odor08} \'Odor G 2008 {\it Universality in non-equilibrium lattice systems}
  (Singapour: World Scientific)

\bibitem{krapivsky10a} Krapivsky P~L, Redner S and Ben-Naim E 2010 {\it A Kinetic View of Statistical
  Physics} (New York: Cambridge University Press) p 414

\bibitem{tauber17} T\"auber U C 2017 {\it Ann. Rev. Cond. Matter Phys.} {\bf 8} 1
  
\bibitem{smoluchowski16} Smoluchowski M 1916 {\it Physik. Z.} {\bf 17} 557

\bibitem{burlatsky87} Burlatsky S F and Ovchinnikov A A 1987 {\it Sov. Phys. JETP} {\bf 65} 908

\bibitem{ovchinnikov78} Ovchinnikov A A and Zeldovich Ya B 1978 {\it Chem. Phys.} {\bf 28} 215

\bibitem{toussaint83} Toussaint D and Wilczek F 1983 {\it J. Chem. Phys.} {\bf 78} 2642

\bibitem{kang84} Kang K and Redner S 1984 {\it Phys. Rev. Lett.} {\bf 52} 955

\bibitem{galfi88} G\'alfi L and R\'acz Z 1988 {\it Phys. Rev. A} {\bf 38} 3151

\bibitem{krapivsky10b} Krapivsky P~L, Redner S and Ben-Naim E 2010 {\it A Kinetic View 
of Statistical Physics} (New York: Cambridge University Press) p 435

\bibitem{bramson88} Bramson M and Lebowitz J L 1988 {\it Phys. Rev. Lett.} {\bf 61} 2397
Erratum 1989 {\it Phys. Rev. Lett.} {\bf 62} 694

\bibitem{bramson91} Bramson M and Lebowitz J L 1991 {\it J. Stat. Phys.} {\bf 65} 941

\bibitem{lee95} Lee B P and Cardy J 1995 {\it J. Stat. Phys.} {\bf 80} 971

\bibitem{tauber05} T\"auber U C, Howard M and Vollmayr-Lee B P 2005 
{\it J. Phys.A: Math. Gen.} {\bf 38} R79

\bibitem{leyvraz92} Leyvraz F and Redner S 1992 {\it Phys. Rev. A} {\bf 46} 3132

\bibitem{benavraham86} Ben-Avraham D and Redner S 1986 {\it Phys. Rev. A} {\bf 34} 501

\bibitem{hilhorst04} Hilhorst H J, Deloubri\`ere O, Washenberger M J and T\"auber U C 
2004 {\it J. Phys.A: Math. Gen.} {\bf 37} 7063

\bibitem{lee84} Lee K and Weinberg E J 1984 {\it Nucl. Phys. B} {\bf 246} 354

\bibitem{ovchinnikov89} Ovchinnikov A A, Timashev S F and Belyi A A 1989 {\it Kinetics 
of Diffusion Controlled Chemical Processes} (Hauppauge: Nova Science)

\bibitem{savara10} Savara A and Weitz E 2010 {\it J. Phys. Chem. C} {\bf 114} 20621

\bibitem{vardeny80} Vardeny Z, O'Connor P, Ray S and Tauc J 1980 {\it Phys. Rev. Lett.} {\bf 44} 1267

\bibitem{turban18} Turban L and Fortin J-Y 2018 {\it J. Phys. A: Math. Theor.} {\bf 51} 145001

\bibitem{ojo2001} Ojo M O 2001 {\it Kragujevac J. Math.} {\bf 23} 101

\bibitem{pinheiro2016} Pinheiro E C and Ferrari S L P 2016 {\it J. Stat. Comp. Sim.} {\bf 86} 2241

\bibitem{riordan79} Riordan J 1979 {\it Combinatorial identities} (Huntington, New York: 
Robert E. Krieger Publishing Company) p 66

\bibitem{gradshteyn80} Gradshteyn I S and Ryzhik I M 1980 {\it Tables of Integrals, 
Series, and Products} (New York: Academic Press) (7.127) p 797

\bibitem{whittaker27} Whittaker E T and Watson G N 1927 {\it A Course of Modern 
Analysis} (Cambridge: Cambridge University Press) p 462 

\bibitem{turban15} Turban L 2015 {\it J. Phys. A: Math. Theor.} {\bf 48} 445001

\bibitem{chupeau15} Chupeau M, B\'enichou O and Voituriez R 2015 {\it Nat. Phys.} 
{\bf 11} 844

\bibitem{foltin94} Foltin G, Oerding K, R\'acz Z, Workman R and Zia R 1994 {\it Phys.
Rev. E} {\bf 50} R639

\bibitem{edwards82} Edwards S F and Wilkinson D R 1982 {\it Proc. Roy. Soc. London Ser. A} 
{\bf 381} 17

\bibitem{majumdar04}
Majumdar S~N and Comtet A 2004 {\em Phys. Rev. Lett.\/} {\bf 92} 225501

\bibitem{majumdar05a}
Majumdar S~N and Comtet A 2005 {\em J. Stat. Phys.\/} {\bf 119} 777

\bibitem{majumdar05b}
Majumdar S~N 2005 {\em Current Science\/} {\bf 89} 2076

\bibitem{fortin15} Fortin J-Y and Clusel M 2015 {\it J. Phys. A: Math. Theor.} {\bf 48} 183001

\end{thebibliography}

\end{document}